\documentclass[12pt]{article}

\usepackage{amssymb}
\usepackage{amsmath}
\usepackage{amsfonts}
\usepackage{amsbsy}
\usepackage{epsfig}
\usepackage{psfrag}

\begin{document}

\title{Breaking scale invariance \\ from a singular inflaton potential}

\author{
Jinn-Ouk Gong\footnote{jgong@muon.kaist.ac.kr} \\
{\em Department of Physics, KAIST, Daejeon, Republic of Korea}}

\date{\today}

\maketitle

\begin{abstract}

In this paper we break the scale invariance of the primordial power spectrum of
curvature perturbations of inflation. Introducing a singular behaviour due to
spontaneous symmetry breaking in the inflaton potential, we obtain fully
analytic expressions of scale dependent oscillation and a modulation in power on
small scale in the primordial spectrum. And we give the associated cosmic
microwave background and matter power spectra which we can observe now and
discuss the signature of the scale dependence. We also address the possibility
of whether some inflationary model with featured potential might mimic the
predictions of the scale invariant power spectrum. We present some examples
which illustrate such degeneracies.

\end{abstract}

\vspace*{-75ex}
\hspace*{\fill} KAIST-TH/2005-06

\thispagestyle{empty}
\setcounter{page}{0}
\newpage
\setcounter{page}{1}

\section{Introduction}
\label{intro}

Inflation \cite{inf}, an epoch of accelerated expansion in the earliest history
of the universe, is believed to be able to solve many problems of standard big
bang cosmology, such as homogeneity, isotropy and flatness of the universe we
observe now. Among them, one of the most important for current and future
observations is the generation of density perturbations from quantum
fluctuations in the inflaton field \cite{pert,paction}. Once those perturbations
develop, they become the seeds of the fluctuations in the otherwise homogeneous
and isotropic cosmic microwave background (CMB) and matter distribution in the
subsequent evolution of the universe.

For inflationary models driven by a scalar field slowly rolling down the
potential, the density perturbations produced during inflation are generally
expected to be nearly scale invariant; that is, the power spectrum of those
primordial perturbations are predicted to be $\mathcal{P}(k) \propto k^{n - 1}$
with $n \sim 1$, where the deviation from $n = 1$ depends on the detailed
dynamics of the inflaton. In this `slow-roll' approximation we have the
spectral index $n$ as a function of the potential and its derivatives, or
slow-roll parameters, which distinguish different inflationary models. They are
required to be small in common to maintain the scale invariance, which is as
stated above, the consequence of the inflaton field slowly rolling down the
smooth potential. Certainly, recent observations put strong bound on $n$; it
must be close to 1 \cite{wmapparameter}.

However, the scale invariance of $\mathcal{P}(k)$ is not directly constrained
by experiments but {\em assumed}. The spectra we can observe are the CMB and
matter power spectra, which are believed to have been generated from the common
$\mathcal{P}(k)$, multiplied by some transfer functions describing the
complicated physics of the evolution. We believe the scale invariance since the
observations on large scales, where the linearity preserves the basic
properties of primordial fluctuations, fit fairly with $\mathcal{P}(k)$ with $n
\sim 1$, not since we actually observe the very scale invariant
$\mathcal{P}(k)$ directly, nor since it is the only theoretical possibility. As
mentioned above, when the inflaton rolls down the smooth potential slowly, the
power spectrum becomes very close to the flat one, with $n \sim 1$. But the
slow-roll approximation does not encompass all the possibilities of inflation.
For example, even the potential exhibits singular behaviours and the inflaton
field does not slowly roll down the potential so that the standard slow-roll
scheme is not applicable \cite{gsr,cgs}, still it is possible to keep inflation
going on. In this case, generally we obtain a scale dependent primordial power
spectrum $\mathcal{P}(k)$ \cite{sbb,starobinsky,kl,shenhancement,ace,hs},
possibly leading to a detectable discrepancy between the standard cosmological
model and observations such as the dearth of halo substructure \cite{kl} and
the suppression of the low multipoles in the CMB anisotropy \cite{cpkl}.

As the origin of such singular behaviour, we can think of the general situation
that inflaton field $\phi$ is coupled to other scalar fields. During inflation,
those scalar fields may undergo spontaneous symmetry breaking, so that they
acquire some nonzero vacuum expectation values. Provided that the symmetry
breaking is rapid, the coupled scalar fields would result in a non-smooth
behaviour in the inflaton potential \cite{hs,minf}. For example, consider a
simple toy model where the effective potential contains quadratic terms of
scalar fields $\phi$ and $\varphi$ as
\begin{equation}
V(\phi,\varphi) = V_0 - \mu^3\phi - m_{\varphi\phi}^2\varphi\phi \, .
\end{equation}
When $\varphi$ acquires vacuum expectation value $v$, we can see that around
the moment of symmetry breaking the `slope' of the potential $V(\phi)$ changes.
For a more realistic example, we move to another simple model where the
effective potential of two scalar fields $\phi$ and $\varphi$ is given as
\cite{sbb}
\begin{equation}
V(\phi,\varphi) = V_0 - \frac{1}{2} m_\phi^2 \phi^2 - \frac{1}{2} g^2 \phi^2
\varphi^2 \, .
\end{equation}
Now assume that by spontaneous symmetry breaking the field $\varphi$ acquires a
nonzero vacuum expectation value $\langle \varphi \rangle = v$ when $\phi$
becomes greater than some critical value $\phi_c$; then the mass of $\phi$ is
modified. Then the corresponding potential is
\begin{equation}
V(\phi) = V_0 - \frac{1}{2} \left( m_\phi^2 + g^2 v^2 \right) \phi^2 \, .
\end{equation}
Clearly we see that there is a downward behaviour at the moment of symmetry
breaking\footnote{As is well known, such a spontaneous symmetry breaking plays a
crucial role in the potential of the hybrid inflation \cite{hybrid}
\begin{equation}
V(\phi,\varphi) = \frac{1}{4\lambda} \left( M^2 - \lambda\varphi^2 \right)^2 +
\frac{1}{2} m^2 \phi^2 + \frac{1}{2} g^2 \phi^2 \varphi^2 \, . \nonumber
\end{equation}
The effective mass of the field $\varphi$ is $-M^2 + g^2\phi^2$, so for $\phi <
\phi_c = M/g$, the minimum of $\varphi$ changes from 0 to $\pm
M/\sqrt{\lambda}$, consequently the mass of $\phi$ also changes and we obtain a
very sharp, step-like downward feature (``waterfall"), depending on the choice
of the parameters. Note that in the usual hybrid inflation models such a
spontaneous symmetry breaking is used to finish the inflationary phase, but in
the present work we are interested in the associated power spectrum provided
that inflation is {\em not suspended} by the symmetry breaking and is still
proceeding.}.

Thus, it is meaningful and interesting to open the possibility of non-smooth,
featured potential and to investigate the resulting power spectrum on both
theoretical and observational grounds. Here we consider some inflaton potentials
with singular, non-smooth behaviours and calculate the fully analytic results
for the primordial power spectrum $\mathcal{P}(k)$ of the curvature
perturbations. Using those $\mathcal{P}(k)$, we present the subsequent CMB and
matter power spectra relevant for observations, and discuss the viability of
such featured potentials. The possibility of broken scale invariance has been
studied several times numerically \cite{sbb,kl,shenhancement,ace}, but here we
obtain fully analytic results for $\mathcal{P}(k)$ as far as we can.

This paper is organised as follows. First, in Section \ref{secgsr} we review the
generalised slow-roll formalism, which we will use in the following Section
\ref{secps} to derive the primordial power spectrum of curvature perturbations
for potentials with feature such as slope change and step. In Section
\ref{secobservation} we describe the CMB and matter power spectra associated
with $\mathcal{P}(k)$ of the previous section, and discuss the possibility of
degeneracy surviving the current observational constraints. In Section
\ref{seccon} we summarise the results and conclude. Throughout the paper we set
$c = \hbar = 8\pi G = 1$.

\section{General Slow-Roll Formalism}
\label{secgsr}

When we deal with some inflaton potential which is not smooth, it is not
guaranteed that the slow-roll parameters,
\begin{equation}
\epsilon \equiv -\frac{\dot H}{H^2} = -\frac{d\ln H}{d\ln a} \hspace{1cm}
\mbox{and} \hspace{1cm} \delta_1 \equiv \frac{\ddot\phi}{H\dot\phi} =
\frac{d\ln\dot\phi}{d\ln a} \, ,
\end{equation}
are approximately constants. Therefore, naively applying the standard slow-roll
approximation, where it is additionally assumed that $\epsilon$ and $\delta_1$
are both slowly varying, does not work. Instead we must use some other scheme
which encompasses the cases where the standard slow-roll picture is not
applicable to solve the situation analytically; we adopt the general slow-roll
picture. In this section, we briefly review the Green's function method
\cite{gs} and the general slow-roll approximation discussed in \cite{gsr,cgs}.
Then we write the power spectrum of the density perturbations which we will use
in the next section.

The scalar perturbation to the homogeneous, isotropic background metric is
generally given by \cite{paction}
\begin{equation}
ds^2 = a^2(\eta) \left\{ (1 + 2A) d\eta^2 - 2\partial_iB dx^id\eta - [ (1 -
2\psi)\delta_{ij} + 2\partial_i\partial_jE] dx^idx^j \right\} \, ,
\end{equation}
and we define
\begin{equation}
\varphi = a \left( \delta\phi + \frac{\dot\phi}{H} \psi \right) \hspace{1cm}
\mbox{and} \hspace{1cm} z = \frac{a\dot\phi}{H} \, ,
\end{equation}
then the intrinsic curvature perturbation of the comoving hypersurfaces is given
by
\begin{equation}
\mathcal{R}_\mathrm{c} = \frac{\varphi}{z} \, ,
\end{equation}
and the equation of motion for the Fourier component of $\varphi$ is given by
\begin{equation}\label{eom}
\frac{d^2\varphi_k}{d\eta^2} + \left( k^2 - \frac{1}{z} \frac{d^2 z}{d\eta^2}
\right) \varphi_k = 0 \, ,
\end{equation}
where $ \varphi_k $ satisfies the boundary conditions
\begin{equation}
\varphi_k \longrightarrow \left\{
\begin{array}{ccc}
\frac{1}{\sqrt{2k}\,} e^{-ik\eta} & \mbox{as} & -k\eta \rightarrow \infty
\\
A_k z & \mbox{as} & -k\eta \rightarrow 0 \, .
\end{array}
\right.
\end{equation}
Now, writing $ y \equiv \sqrt{2k} \, \varphi_k $, $ x \equiv -k \eta$ and
\begin{equation}
f(\ln x) \equiv \frac{2\pi x z}{k} = \frac{2\pi}{H} \frac{ax\dot\phi}{k} \, ,
\end{equation}
Eq.~(\ref{eom}) becomes
\begin{equation}\label{eom2}
\frac{d^2y}{dx^2} + \left( 1 - \frac{2}{x^2} \right)y = \frac{g(x)}{x^2} y \, ,
\end{equation}
where $g \equiv (f'' - 3f')/f$ and $f' \equiv df/d\ln x$. Using Green's function
method, we can present the solution of Eq.~(\ref{eom2}) as an integral equation
\begin{eqnarray}\label{sol}
y(x) & = & y_0(x) + \frac{i}{2} \int_{x}^{\infty} \frac{du}{u^2} \, g(u) \left[
y_0^*(u) \, y_0(x) - y_0^*(x) \, y_0(u) \right] y(u)
\nonumber \\
& \equiv & y_0(x) + L(x,u) \, y(u) \, ,
\end{eqnarray}
where
\begin{equation}\label{hsol}
y_0(x) = \left(1 + \frac{i}{x}\right) e^{ix}
\end{equation}
is the homogeneous solution with desired asymptotic behaviour.

The power spectrum for the curvature perturbation $\mathcal{P}(k)$ is defined by
\begin{equation}
\frac{2\pi^2}{k^3}\,\mathcal{P}(k)\,\delta^{(3)}(\mathbf{k-l}) = \langle
\mathcal{R}_\mathrm{c}(\mathbf{k})\,
{\mathcal{R}_\mathrm{c}}^\dagger(\mathbf{l}) \rangle \, ,
\end{equation}
which, using the above results, we can write conveniently as
\begin{equation}\label{psy}
\mathcal{P}(k) = \lim_{x \rightarrow 0} \left| \frac{xy}{f} \right|^2 \, .
\end{equation}
Here, we {\em only} assume that $y(x)$ is given approximately by the scale
invariant $y_0(x)$, or equivalently that $g$ is small. It is very important to
note that this is the only assumption we make as required by observations. Then,
since we are interested in the second order corrections, we iterate
Eq.~(\ref{sol}) twice, i.e.,
\begin{equation}\label{yx2}
y(x) \simeq y_0(x) + L(x,u) \, y_0(u) + L(x,u) \, L(u,v) \, y_0(v) \, .
\end{equation}
Substituting into Eq.~(\ref{psy}), and skipping over calculations, we can
rewrite the power spectrum as \cite{cgs}
\begin{eqnarray}\label{spectrum1}
\ln \mathcal{P}(k) & = & \ln \left( \frac{1}{f_\star^2} \right) + \frac{2}{3}
\frac{f'_\star}{f_\star} + \frac{1}{9} \left( \frac{f'_\star}{f_\star} \right)^2
+ \frac{2}{3} \int_0^\infty \frac{du}{u} \, W_\theta(x_\star,u) \, g(u) +
\frac{2}{9} \left[ \int_0^\infty \frac{du}{u} \, X(u) \, g(u) \right]^2
\nonumber \\
& & \mbox{} - \frac{2}{3}\int_0^\infty \frac{du}{u} \, X(u) \, g(u)
\int_u^\infty \frac{dv}{v^2} \, g(v) - \frac{2}{3} \int_0^\infty \frac{du}{u} \,
X_\theta(x_\star,u) \, g(u) \int_u^\infty \frac{dv}{v^4} \, g(v)
\\
\label{spectrum2}
& = & \ln \left(\frac{1}{f_\star^2}\right) - 2 \int_0^\infty \frac{du}{u} \,
w_\theta(x_\star,u) \, \frac{f'}{f} + 2 \left[ \int_0^\infty \frac{du}{u} \,
\chi(u) \, \frac{f'}{f} \right]^2 \nonumber \\
& & \mbox{} - 4 \int_0^\infty \frac{du}{u} \, \chi(u) \, \frac{f'}{f}
\int_u^\infty \frac{dv}{v^2} \, \frac{f'}{f}
\end{eqnarray}
where the window functions are given as
\begin{eqnarray}
W(x) & = & \frac{3\sin(2x)}{2x^3} - \frac{3\cos(2x)}{x^2} - \frac{3\sin(2x)}{2x}
\, ,
\nonumber \\
X(x) & = & -\frac{3\cos(2x)}{2x^3} - \frac{3\sin(2x)}{x^2} +
\frac{3\cos(2x)}{2x} + \frac{3(1 + x^2)}{2x^3} \, ,
\nonumber \\
W_\theta(x_\star,x) & = & W(x) - \theta(x_\star-x) \, ,
\nonumber \\
X_\theta(x_\star,x) & = & X(x) - \frac{x^3}{3} \, \theta(x_\star-x) \, ,
\nonumber \\
w(x) & = & W(x) + \frac{x}{3}W'(x) \, ,
\nonumber \\
\chi(x) & = & X(x) + \frac{x}{3}X'(x)\, ,
\nonumber \\
w_\theta(x_\star,x) & = & w(x) - \theta(x_\star-x) \, ,
\end{eqnarray}
and the subscript $\star$ means some convenient point of evaluation around
horizon crossing.

\section{$\mathcal{P}(k)$ from non-smooth inflaton potential}
\label{secps}

Now, using Eqs.~(\ref{spectrum1}) and (\ref{spectrum2}) for the power spectrum
derived in the previous section, we consider some curvature power spectra
$\mathcal{P}(k)$ for the models of inflation arising from the potentials with
some singular behaviours.

Before we start this section, let us remind that we would like to consider the
cases where we {\em cannot} apply the slow-roll conditions usually assumed;
$\dot\phi$ or $\ddot\phi$ is changing drastically, and the features should be
extremely singular, or very sharp. Then, the somewhat ad hoc features we are
going to discuss here are good enough approximations, and they are our starting
point. Also note that since mild enough features are highly dependent on the
detail of specific models under consideration, it is not easy to extract other
than qualitative behaviour. But sharp cases are universal; as we will see, we
need only a single or a couple of parameters to characterise the feature, and
we acquire complete control of that singularity.

\subsection{Linear potential with a slope change}
\label{slopechange}

This kind of model has been investigated several times \cite{cgs,starobinsky},
so we skip straightforward calculations and just present the situation and the
result. We consider an inflaton $\phi$ rolling down a linear potential with
slope changing from $-A$ to $- A - \Delta A$ at $\phi = \phi_0$. We can write
the potential as
\begin{equation}
V(\phi) = V_0 \left\{ 1 - \left[ A + \theta(\phi-\phi_0) \, \Delta A \right]
\left( \phi - \phi_0 \right) \right\} \, .
\end{equation}
Assuming $|A| \ll 1$ so that $3 H^2 = V_0$ and $x = k/(aH)$, and solving the
equation of motion for $\phi$, we obtain
\begin{equation}\label{slopechangedphidN}
\frac{d\phi}{dN} = A + \theta(N-N_0) \, \Delta A \left[ 1 - e^{-3(N-N_0)}
\right] \, ,
\end{equation}
where $N = \int H \, dt$ and $\phi(N_0) = \phi_0$. Then if $|\Delta A / A| \ll
1$ so that the approximate scale invariance of the spectrum is maintained, we
have
\begin{equation}\label{v''}
\frac{V''}{V} = -\frac{1}{3}g = - \Delta A \, \delta(\phi-\phi_0) = -
\frac{\Delta A}{A} \, \delta(N-N_0) \, ,
\end{equation}
where $x_0 = k/(aH)_0 = k/k_0$. Substituting into Eq.~(\ref{spectrum1}) and
performing the integration, we find the power spectrum as \cite{cgs}
\begin{eqnarray}\label{slopechangelnP}
\ln \mathcal{P} & = & \ln \left( \frac{V_0}{12\pi^2A^2} \right) + 2\left(
\frac{\Delta A}{A} \right) \left[ W(x_0) - 1 \right]
\nonumber \\
& & + 2 \left( \frac{\Delta A}{A} \right)^2 \left\{ X(x_0) \left[ X(x_0) -
\frac{3(1 + x_0^2)}{2x_0^3} \right] + \frac{1}{2} \right\} \, .
\end{eqnarray}
The spectrum $\ln\mathcal{P}$ is plotted in the left panel of
Figure~\ref{slopechange}. In addition to the constant leading term $\ln \left[
V_0/(12\pi^2A^2) \right]$, the most important change from the scale invariant
$\mathcal{P}(k)$ is that there is a modulation in power, a suppression (an
enhancement), after the slope change when the change is positive (negative).
Also we obtain an diminishing oscillation after the slope change. Another
noteworthy point is the behaviour of $d\phi/dN$ shown in the right panel of
Figure~\ref{slopechange}. When $\Delta A < 0$, i.e., the slope becomes flatter,
$d\phi/dN$ gets smaller than its value at horizon crossing. Then, the
perturbations on the scales smaller than the very scale of slope change
experience enhancement after they go outside the horizon\footnote{I am grateful
to Misao Sasaki for explaining these ideas.}; as shown in the left panel of
Figure~\ref{slopechange}, $\mathcal{P}(k)$ is rising across $\ln x_0 = 0$. This
agrees precisely with Ref.~\cite{shenhancement}.

\begin{figure}[h]
\begin{center}
\epsfig{file=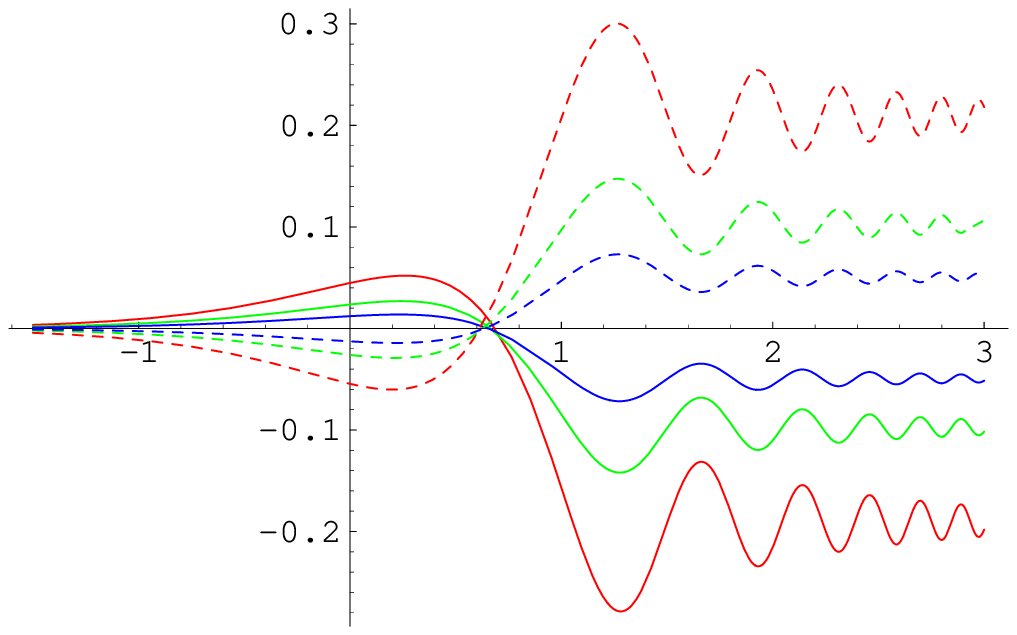, width = 8.1cm}%
\epsfig{file=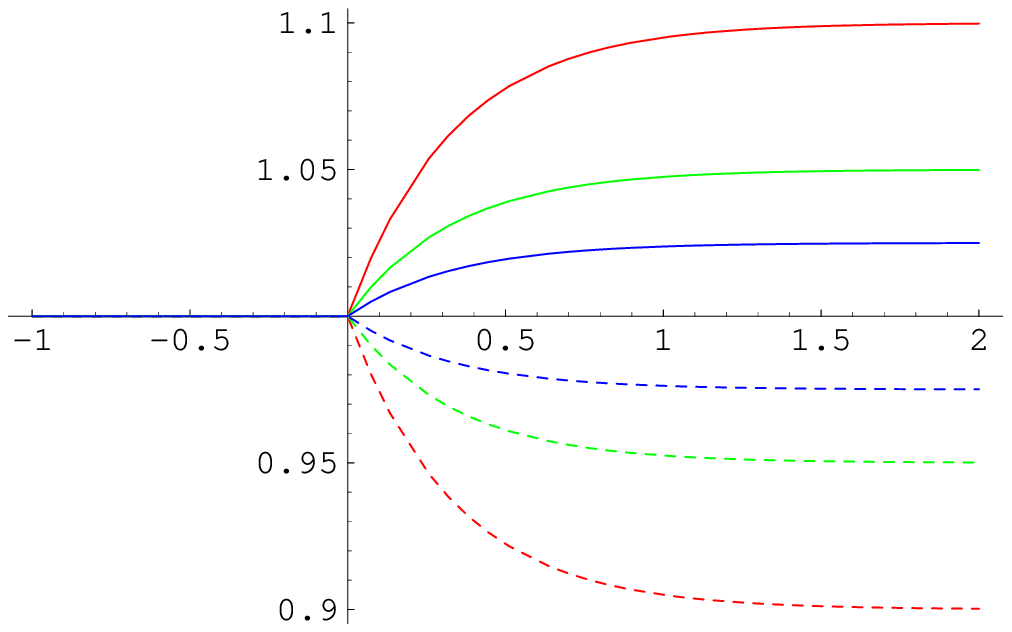, width = 8.1cm}%
\caption{\label{slopechange}(Left) plot of $\ln\mathcal{P}$ versus $\ln x_0$,
Eq.~(\ref{slopechangelnP}), and (right) $d\phi/dN$ versus $N$,
Eq.~(\ref{slopechangedphidN}), where $A$ is normalised to 1. Solid lines
correspond to $\Delta A/A > 0$ and dashed lines to $\Delta A/A < 0$. $\left|
\Delta A/A \right|$ are set to 0.025, 0.05 and 0.1 from the innermost line. As
expected, the modulation in power becomes stronger as the slope change is
steeper.}
\end{center}
\end{figure}

\subsection{Linear potential with a sharp downward step}
\label{lpstep}

Now, we consider the situation where the potential is linear with slope $-A$,
and there is a sharp downward step of size $aV_0$ at $\phi_s$\footnote{This
model was briefly discussed in Ref.~\cite{inv}. Also some analytic results
regarding this case are given in Ref.~\cite{starobinsky}.}. The potential can be
written as
\begin{equation}
V(\phi) = V_0 \left[ 1 - A (\phi - \phi_s) - a \theta(\phi - \phi_s) \right] \,
,
\end{equation}
where we assume that $A \ll 1$, $a \ll 1$ and $a \ll A^2$ so that inflation is
not suspended in spite of the existence of the step and $3 H^2 = V_0$.

\begin{figure}[h]
\centerline{\epsfxsize=4.5 in \epsfbox{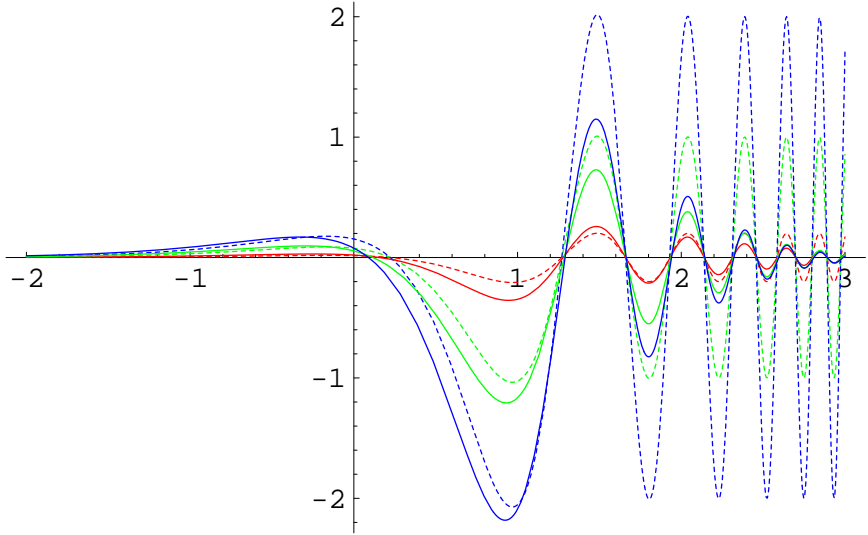}}
\caption{\label{stepgraph} (Dashed) plot of $\ln\mathcal{P}$ versus $\ln x_s$
for a sharp step, Eq.~(\ref{lpsspec}), and (solid) an arctangent step,
Eq.~(\ref{arctanlogp}). The parameters for the arctangent step are set to {$a =
0.001$, $b = 0.02 \times (0.8)^n$ $A = (0.8)^{3n + 1}$} where $n = 2, 3$ and
7/2 from the innermost line. These numbers are chosen to roughly match the
first few peaks of the sharp step cases with $B = 0.1, 0.5$ and 1.0.}
\end{figure}

In this case, we are unable to proceed as before because of the singularity in
the slope, i.e., we have a delta function for $V'(\phi)$. Rather, we use the
relation $f'/f = -\phi''/\phi'$, where $\phi' = d\phi/dN$, and
Eq.~(\ref{spectrum2}). The equation of motion for $\phi$ is
\begin{equation}
\phi'' + 3\phi' - 3A = 3a \delta(\phi - \phi_s) \, ,
\end{equation}
where $\phi_s = \phi(N_s)$, so we have two solutions
\begin{equation}
\phi(N) = \left\{ \begin{array}{lll} A(N - N_s) & \equiv \phi_1(N), &
\mbox{before the step} \\ A\left[ N - e^{-3(N - N_s)}N_s \right] & \equiv
\phi_2(N), & \mbox{after the step} \end{array} \right.
\end{equation}
where we set $\phi_s = 0$. Then, we can write the general solution of $\phi(N)$
as
\begin{equation}
\phi(N) = \phi_1(N) - \frac{B}{3N_s} \left[ \phi_1(N) - \phi_2(N) \right]
\theta(N - N_s)
\end{equation}
where we introduced a new parameter $B \equiv aV_0/A^2 \ll 1$, which corresponds
to the magnitude of the step. Then, we can write
\begin{eqnarray}
\frac{f'}{f} & = &  -\frac{\phi''}{\phi'}
\nonumber \\
& = & -B \left[ \delta(x - x_s) - 3\theta(x_s - x) \right] \left( \frac{x}{x_s}
\right)^3 + B^2 \left[ \theta(x_s - x) \delta(x - x_s) - 3\theta(x_s - x)
\right] \left( \frac{x}{x_s} \right)^6 \, ,
\nonumber \\
\end{eqnarray}
so from Eq.~(\ref{spectrum2}) we obtain the power spectrum as
\begin{eqnarray}\label{lpsspec}
\ln\mathcal{P} & = & \ln \left( \frac{V_0}{12\pi^2A^2} \right) + B \left[ -
\frac{3\sin(2x_s)}{x_s^3} + \frac{6\cos(2x_s)}{x_s^2} + \frac{5\sin(2x_s)}{x_s}
- 2\cos(2x_s) \right]
\nonumber \\
& & + B^2 \left[ 1+ \frac{9}{4x_s^6} + \frac{3}{2x_s^4} + \frac{1}{4x_s^2} -
\frac{9\cos(2x_s)}{2x_s^6} - \frac{9\sin(2x_s)}{x_s^5} +
\frac{6\cos(2x_s)}{x_s^4} + \frac{5\cos(2x_s)}{2x_s^2} \right.
\nonumber \\
& & \hspace{1.5cm} + \cos(2x_s) + \frac{9\cos(4x_s)}{4x_s^6} +
\frac{9\sin(4x_s)}{x_s^5} - \frac{33\cos(4x_s)}{2x_s^4} -
\frac{18\sin(4x_s)}{x_s^3}
\nonumber \\
& & \hspace{1.5cm} \left. + \frac{49\cos(4x_s)}{4x_s^2} +
\frac{5\sin(4x_s)}{x_s} - \cos(4x_s) \right] \, .
\end{eqnarray}
Interestingly, now we have a scale dependent oscillatory behaviour in the
spectrum. The spectra, Eq.~(\ref{lpsspec}), with a few different values of $B$
are shown in Figure~\ref{stepgraph}. Across the step, $\ln(k/k_s) = 0$, there is
an oscillation the amplitude of which depends on $B$. There are two important
differences from the case of a slope change. One thing is that there is no power
modulation in the small scale on average; the power spectrum oscillates around
the value without the step. The other important point is that the oscillation
does not diminish, and survives down to the small scales. These two points
become distinct compared with the case of slope change in the CMB and matter
power spectra which shall be discussed in the following section.

\subsection{Linear potential with 2 successive slope changes}
\label{2slopechanges}

\begin{figure}[h]
\centerline{\epsfxsize=4.5 in \epsfbox{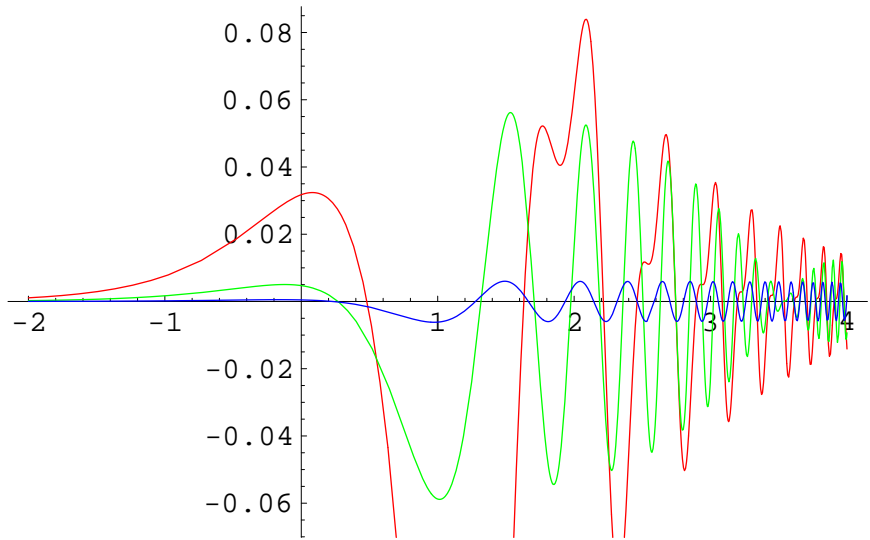}}
\caption{\label{2slopegraph} Plot of $\ln\mathcal{P}$ versus $\ln x_1$ for 2
slope changes, Eq.~(\ref{2slopelogP}). Here $\Delta A/A = 0.1$, and $\alpha =
0.01, 0.1$ and 1.0 from the innermost line.}
\end{figure}

Here, we consider a linear potential with two successive slope changes, from
$-A$ to $-A - \Delta A$ at $\phi_1 = \phi(N_1)$, and then back to $-A$ at
$\phi_2 = \phi(N_2)$. Then, we can write the potential as
\begin{equation}
V(\phi) = V_0 \left\{ 1 - A (\phi - \phi_1) - \Delta A \left[ \theta(\phi -
\phi_1) (\phi - \phi_1) - \theta(\phi - \phi_2) (\phi - \phi_2) \right] \right\}
\, .
\end{equation}
As before, assuming $|A| \ll 1$, we have
\begin{equation}
\frac{V''}{V} = - \frac{\Delta A}{\phi'(N)} \left[ \delta(N - N_1) \theta(N_2 -
N) - \theta(N - N_1) \delta(N_2 - N) \right] \, ,
\end{equation}
and
\begin{equation}
\phi'(N) = A + \theta(N - N_1) \left[ 1 - e^{-3 (N - N_1)} \right] \Delta A -
\theta(N - N_2) \left[ 1 - e^{-3 (N - N_2)} \right] \Delta A \, .
\end{equation}
Now, remembering that $x = k/(aH)$, we write the difference of the position of
two slope changes as
\begin{equation}
N_2 - N_1 = - \ln x_2 + \ln x_1 = \ln (1 + \alpha)
\end{equation}
with $x_2 = x_1/(1 + \alpha)$. Then, first expand only in terms of $\Delta A/A$,
we obtain the power spectrum as
\begin{eqnarray}\label{2slopelogP}
\ln\mathcal{P} & = & \ln\left( \frac{V_0}{12\pi^2A^2} \right) + \frac{\Delta
A}{A} \left\{ \frac{3\sin(2x_1)}{x_1^3} - \frac{3(1 + \alpha)^3\sin\left[2x_1/(1
+ \alpha)\right]}{x_1^3} \right.
\nonumber \\
& & \hspace{4cm} -\frac{6\cos(2x_1)}{x_1^2} + \frac{6(1 +
\alpha)^2\cos\left[2x_1/(1 + \alpha)\right]}{x_1^2}
\nonumber \\
& & \left. \hspace{4cm} - \frac{3\sin(2x_1)}{x_1}  + \frac{3(1 +
\alpha)^3\sin\left[2x_1/(1 + \alpha)\right]}{x_1} \right\} + \mathcal{O} \left[
\left(\frac{\Delta A}{A}\right)^2 \right] \, ,
\nonumber \\
\end{eqnarray}
where the term $\mathcal{O} \left[ \left(\Delta A/A\right)^2 \right]$ is very
long and much more complex so we do not present it. In
Figure~\ref{2slopegraph}, we plot Eq.~(\ref{2slopelogP}). The crucial
difference from the single slope change case is that now there is no change in
power, but {\em diminishing} oscillation back to the value without slope
changes. But as shown in Figure~\ref{2slopegraph}, when $\alpha$ becomes small,
i.e., the slope changes occur very close to each other, the oscillatory
behaviour hardly dies. This becomes very clear if we take the limit that
$\alpha \rightarrow 0$ but the product $\alpha (\Delta A/A)$ remains finite, so
that any excessive power in $\alpha$ alone in the perturbative expansion is
negligible. Note that this is equivalent to the sharp step which corresponds to
the limit $\alpha \rightarrow 0$. Then, introducing a new parameter $B \equiv 3
\alpha (\Delta A/A)$, which should correspond to the height of the step, we
obtain precisely the same power spectrum as the sharp step case,
Eq.~(\ref{lpsspec}). One thing to note is that when the separation is not
negligible, the situation is similar to the smoothed step case. As will be
discussed in detail later, when the step is not infinitely sharp but mild, we
do observe the oscillation diminish (See Figure~\ref{stepgraph}).

\subsection{Inverted square potential with a sharp downward step}
\label{issp}

Next we consider a little more complicated model where inflation occurs while
the field rolls down an inverted square potential with a sharp downward step at
$\phi_s$ of size $aV_0$. Then, the potential is given as
\begin{equation}
V(\phi) = V_0 \left[ 1 - \frac{1}{2}\mu^2\phi^2 - a\theta(\phi - \phi_s) \right]
\, ,
\end{equation}
and as before, we assume that $\mu^2 \ll 1$, $a \ll 1$ and $a \ll \mu^4$. Then,
we can solve the equation of motion for $\phi(N)$ as
\begin{equation}
\phi(N) = \left\{ \begin{array}{lll} e^{r(N - N_s)} & \equiv \phi_+(N), &
\mbox{growing mode} \\
e^{-(3 + r) (N - N_s)} & \equiv \phi_-(N), & \mbox{decaying mode} \end{array}
\right.
\end{equation}
where
\begin{equation}
r = \frac{3}{2} \left( \sqrt{1 + \frac{4}{3}\mu^2} - 1 \right) \, .
\end{equation}
Therefore the general solution can be written as
\begin{equation}
\phi(N) = \phi_+(N) + \frac{\mu^2 B}{3 + 2r} \left[ \phi_+(N) - \phi_-(N)
\right] \theta(N - N_s) \, ,
\end{equation}
where we assume that before the step $\phi(N)$ is completely dominated by the
growing solution only. We also introduced a small parameter $B \equiv aV_0/\mu^4
\ll 1$. Then, we obtain
\begin{eqnarray}
\frac{f'}{f} & = & -\mu^2 + \frac{1}{3}\mu^4 - B \left[\delta(x - x_s) - 3
\theta(x_s - x) \right]\left( \frac{x}{x_s} \right)^3
\nonumber \\
& & - \mu^2B \left[ \frac{1}{3} \delta(x - x_s) + 2 \left( \frac{x}{x_s}
\right)^3 \ln \left( \frac{x}{x_s} \right) \delta(x - x_s) \right.
\nonumber \\
& & \left. \hspace{1.5cm} - 2 \left( \frac{x}{x_s} \right)^3 \theta(x_s - x) - 6
\left( \frac{x}{x_s} \right)^3 \ln \left( \frac{x}{x_s} \right) \theta(x_s - x)
\right]
\nonumber \\
& & - B^2 \left[ - \delta(x - x_s) \theta(x_s - x) + 3 \theta(x_s - x)
\right]\left( \frac{x}{x_s} \right)^6 \, ,
\end{eqnarray}
so that skipping over tedious calculations we finally have the spectrum as
\begin{eqnarray}\label{ispslogP}
\ln\mathcal{P} & = & \ln \left[ \left( \frac{H}{2\pi} \right)^2 \left(
\frac{H}{\dot\phi_\star} \right)^2 \right] + 2\alpha_\star\mu^2 + \left(
-\frac{2}{3}\alpha_\star - 4 + \frac{\pi^2}{2} \right) \mu^4
\nonumber \\
& & + B \left[ -\frac{3\sin(2x_s)}{x_s^3} + \frac{6\cos(2x_s)}{x_s^2} +
\frac{5\sin(2x_s)}{x_s} - 2\cos(2x_s) \right]
\nonumber \\
& & + \mu^2B \left[ -\frac{2}{3} - \frac{3\pi}{x_s^3} - \frac{\pi}{x_s} +
\frac{3\pi\cos(2x_s)}{x_s^3} - \frac{14\sin(2x_s)}{x_s^3} +
\frac{16\cos(2x_s)}{x_s^2} + \frac{6\pi\sin(2x_s)}{x_s^2} \right.
\nonumber \\
& & \hspace{1.5cm} \left.- \frac{5\pi\cos(2x_s)}{x_s} +
\frac{20\sin(2x_s)}{3x_s} - \frac{2}{3}\cos(2x_s) - 2\pi\sin(2x_s) +
\frac{2Si(2x_s)}{x_s} \left( 1 + \frac{3}{x_s^2} \right) \right]
\nonumber \\
& & + B^2 \left[ 1+ \frac{9}{4x_s^6} + \frac{3}{2x_s^4} + \frac{1}{4x_s^2} -
\frac{9\cos(2x_s)}{2x_s^6} - \frac{9\sin(2x_s)}{x_s^5} +
\frac{6\cos(2x_s)}{x_s^4} + \frac{5\cos(2x_s)}{2x_s^2} \right.
\nonumber \\
& & \hspace{1.5cm} + \cos(2x_s) + \frac{9\cos(4x_s)}{4x_s^6} +
\frac{9\sin(4x_s)}{x_s^5} - \frac{33\cos(4x_s)}{2x_s^4} -
\frac{18\sin(4x_s)}{x_s^3}
\nonumber \\
& & \hspace{1.5cm} \left. + \frac{49\cos(4x_s)}{4x_s^2} +
\frac{5\sin(4x_s)}{x_s} - \cos(4x_s) \right] \, .
\end{eqnarray}
The first 3 terms are exactly the power spectrum of inverse square potential
without the step. Again, the $B$ and $B^2$ terms are the same as the previous
results. This is clear from paying our attention to the meaning of $B$; they
are purely due to the existence of the step, so those terms should be the same.
Note that there exists another small parameter, $\mu^2$, which also controls
the shape of $\mathcal{P}(k)$, so we have the combined effect of $\mu^2$ and
$B$, the $\mu^2B$ term.

\subsection{Linear potential with a smoothed downward step}
\label{smoothedstep}

Finally, we present the case of a smoothed step. It is rather formidable to
fully proceed analytically, so here we are satisfied with the integral
equation. For calculational simplicity, we consider a linear potential with a
smoothed step. The step may be described in many different ways, for example,
using an arctangent step:
\begin{equation}\label{arctan}
V(\phi) = V_0 \left[ 1 - A\phi - a\tan^{-1} \left( \frac{\phi - \phi_s}{b}
\right) \right] \, ,
\end{equation}
or a hyperbolic tangent step \cite{ace}:
\begin{equation}\label{tanh}
V(\phi) = V_0 \left[ 1 - A\phi - a\tanh \left( \frac{\phi - \phi_s}{b} \right)
\right] \, ,
\end{equation}
where $a$ and $b$ determine the height and the steepness of the step,
respectively. For example, for the arctangent step, Eq.~(\ref{arctan}), assuming
that $A$, $a$ and $b$ are all small, it is not difficult to proceed the
numerical calculation to obtain the power spectrum. We write $V''/V$ as
\begin{equation}
\frac{V''}{V} = \frac{2aA}{b^3} \frac{\ln x_s - \ln x}{\left[ 1 + (A/b)^2(\ln
x_s - \ln x)^2 \right]^2} \, ,
\end{equation}
substitute this into Eq.~(\ref{spectrum1}) and then integrate. Interestingly, by
extracting out all the $x_\star$ dependence we obtain the constant leading term,
$\ln \left[ V_0/(12\pi^2A^2) \right]$, and the result is independent of
$x_\star$. Writing Eq.~(\ref{spectrum1}) using $V''/V$, we have
\begin{eqnarray}\label{arctanlogp}
\ln \mathcal{P} & = & \ln \left( \frac{V_\star^3}{12\pi^2{V'_\star}^2} \right) -
2 \int_0^\infty \frac{du}{u} \, W_\theta(x_\star,u) \frac{V''}{V} + \cdots
\nonumber \\
& = & \ln \left( \frac{V_0}{12\pi^2A^2} \right) - \frac{4aA}{b^3} \int_0^\infty
\frac{du}{u} W(u) \frac{\ln x_s - \ln u}{\left[ 1 + (A/b)^2 (\ln x_s - \ln u)^2
\right]^2} + \cdots \, ,
\end{eqnarray}
where only first two terms are written for simplicity. In
Figure~\ref{stepgraph}, we plot the power spectra of arctangent step,
Eq.~(\ref{arctanlogp}), enclosed in the sharp step cases. As opposed to the
case of sharp step, the oscillatory behaviour diminishes after the step,
finally returning to the original spectrum. Note that, as discussed before,
this behaviour is reminiscent of the case of two slope changes. We can obtain
similar oscillatory, decreasing graph by solving Eq.~(\ref{tanh}).

\subsection{Summary}
\label{logPsummary}

Before closing this section, let us summarise what we have obtained in this
section. When the inflaton potential $V(\phi)$ is not smooth and exhibits some
singular behaviour as another scalar field coupled to $\phi$ acquires nonzero
vacuum expectation value, it should be either the potential suddenly becomes
flatter or steeper at some point, or it quickly drops by some amount small
enough not to bother inflation. Then the standard slow-roll approximation, which
is an useful and convenient scheme {\em provided that} $V(\phi)$ is both flat
and smooth enough, is not applicable. In this section we simplified these
situations and calculated the corresponding power spectrum $\mathcal{P}(k)$ of
curvature perturbations using the generalised slow-roll picture. The results are
Eqs.~(\ref{slopechangelnP}), (\ref{lpsspec}), (\ref{2slopelogP}),
(\ref{ispslogP}) and (\ref{arctanlogp}).

The crucial difference of these results from the scale invariant
$\mathcal{P}(k)$ is that now it is no more flat and featureless, but generally
shows a scale dependent oscillation. If $V(\phi)$ becomes flatter or steeper,
there is a corresponding modulation in power. But this overall power change does
not exist if the value of $V(\phi)$ quickly drops so that it looks like a sharp
step, and the oscillation hardly diminishes. If the decrease is smoothed out and
not infinitely sharp, the oscillation does die away in contrast.

The two parameters which determine the departure from scale invariance are the
magnitude (e.g., $\Delta A/A, B$) and the position (e.g., $x_0, x_s$) of the
feature. The consequence of varying these two parameters on the corresponding
CMB and matter power spectra we can observe now will be discussed in the
following section.

\section{Observational constraints}
\label{secobservation}

\subsection{The CMB and matter power spectra}
\label{cmbmatter}

In this section, we discuss how the CMB and matter power spectra associated
with those $\mathcal{P}(k)$ would appear when we have the primordial power
spectra $\mathcal{P}(k)$ of the previous section. We use CMBFAST code
\cite{cmbfast} to calculate the CMB anisotropy power spectrum and the matter
transfer function. As our background, we adopt $\Lambda$CDM consistent with
recent observations; we set $\Omega_\mathrm{B} = 0.046, \Omega_\mathrm{DM} =
0.224, \Omega_\Lambda = 0.730$ and $h = 0.72$.

\begin{figure}[h]
\psfrag{LCDM}{$\Lambda$CDM}%
\psfrag{l(l+1)Cl/2p(mK2)}{$l(l + 1)C_l/2\pi (\mu K^2)$}%
\psfrag{l}{$l$}%
\psfrag{P(k)(Mpc)3}{$P(k)(\mathrm{Mpc})^3$}%
\psfrag{k(Mpc-1)}{$k (\mathrm{Mpc}^{-1})$}%
\psfrag{A = 0.1}{$A = 0.1$}%
\psfrag{A = 0.5}{$A = 0.5$}%
\psfrag{A = 1.0}{$A = 1.0$}%
\psfrag{A = -0.1}{$A = -0.1$}%
\psfrag{A = -0.5}{$A = -0.5$}%
\psfrag{A = -1.0}{$A = -1.0$}%
\begin{center}
\epsfig{file=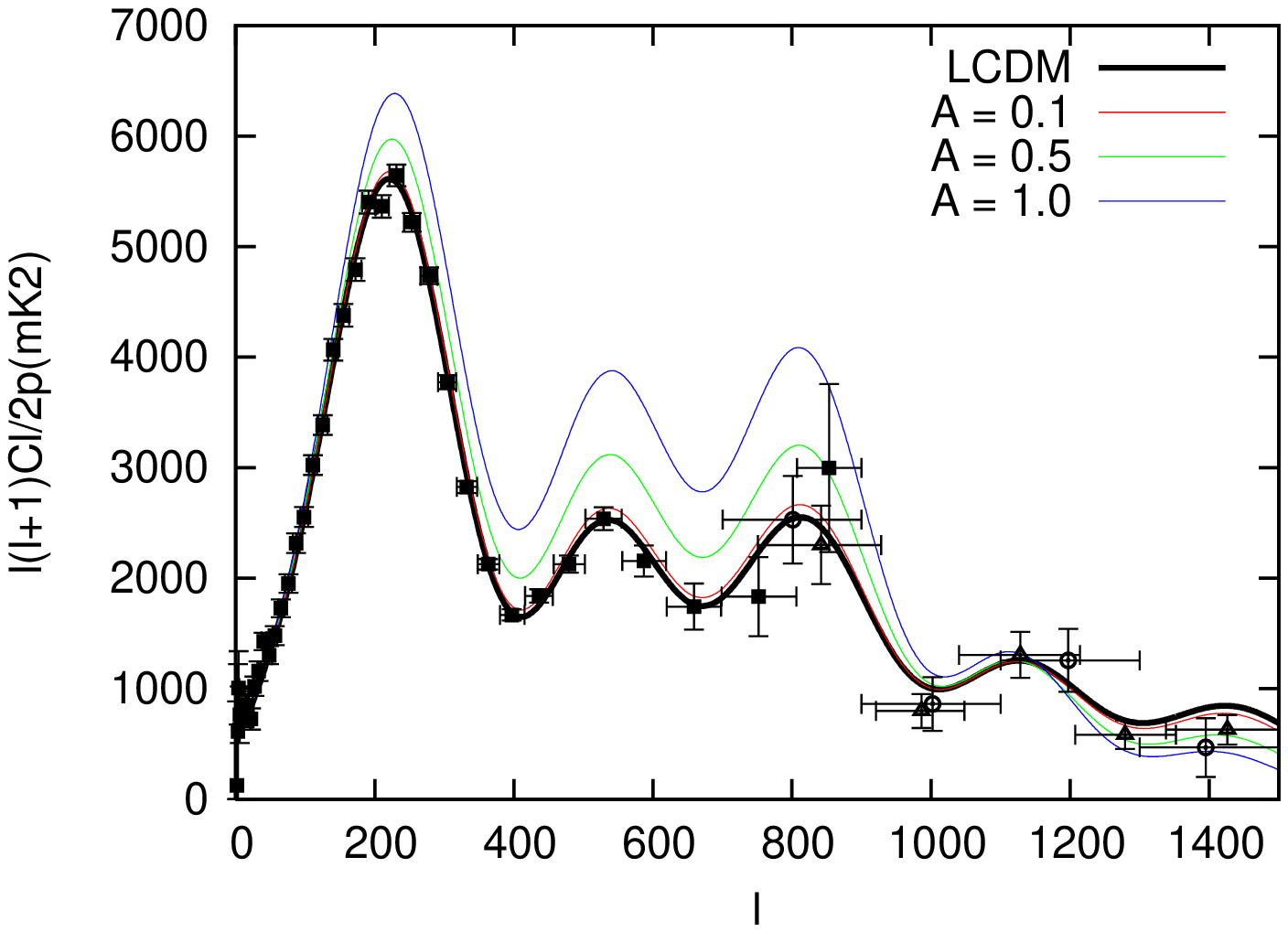, angle = -90, width = 8.1cm}
\epsfig{file=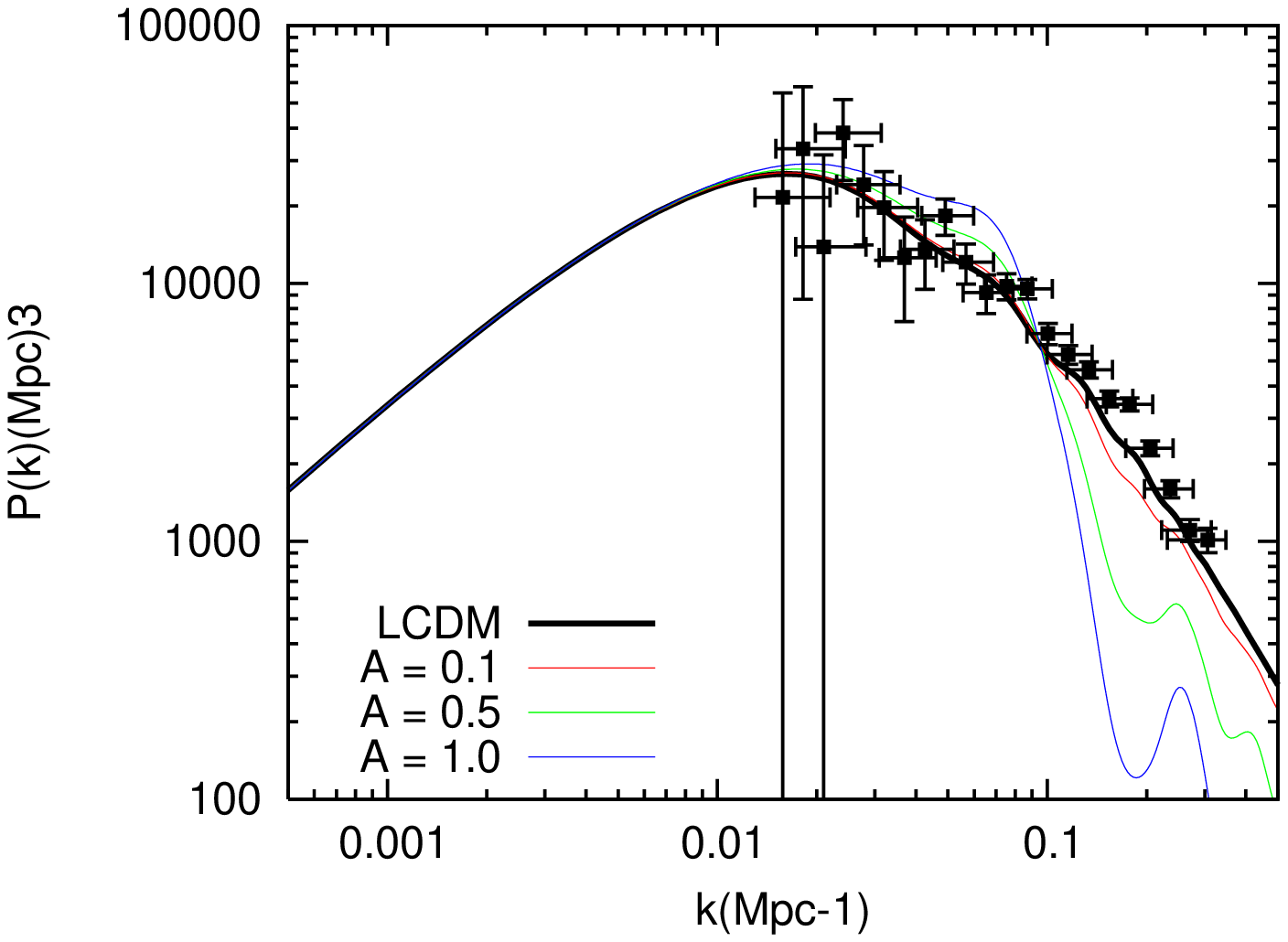, angle = -90, width = 8.1cm}
\end{center}
\begin{center}
\epsfig{file=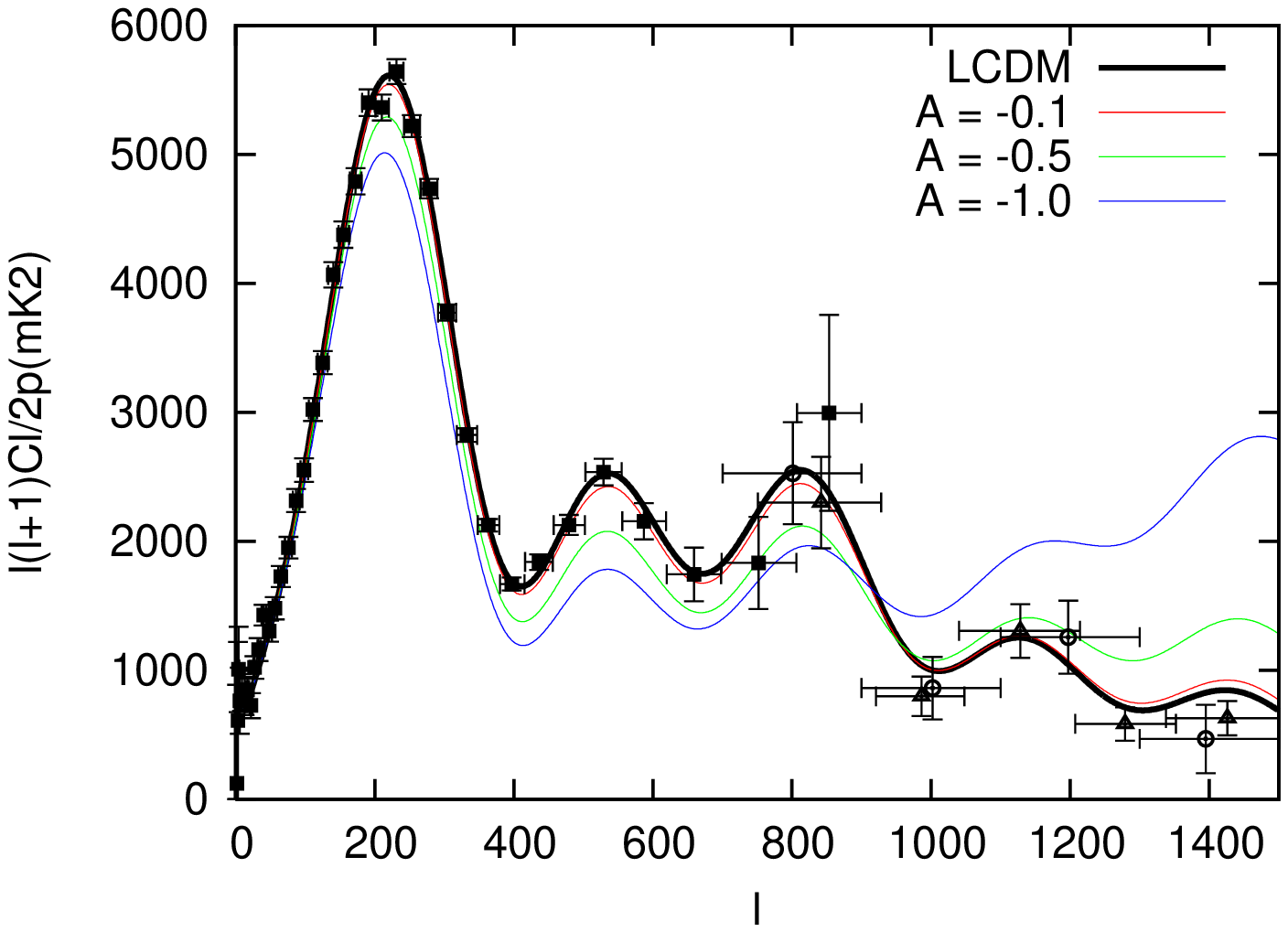, angle = -90, width = 8.1cm}%
\epsfig{file=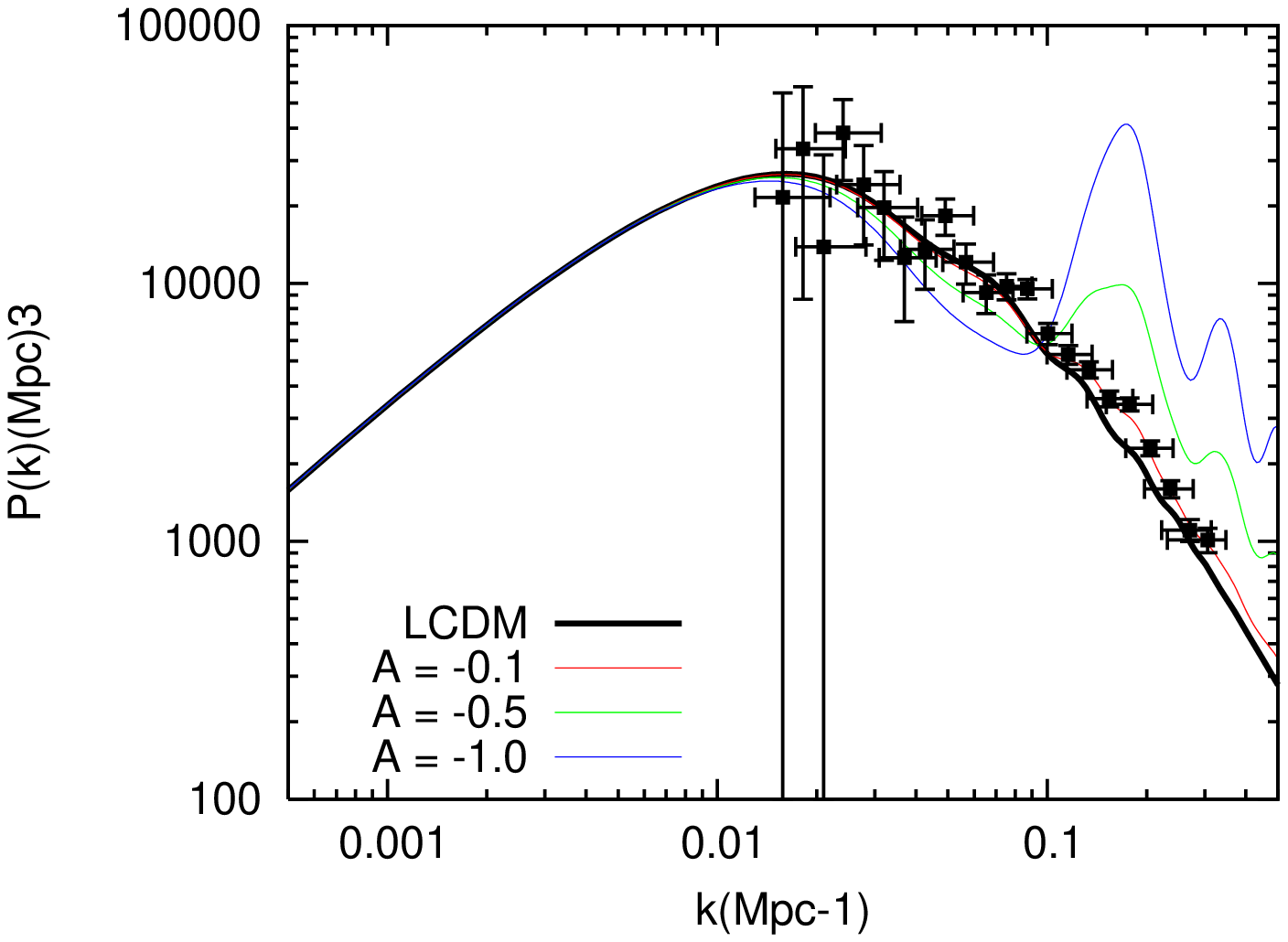, angle = -90, width = 8.1cm}%
\end{center}
\caption{\label{k_size} (Left column) plots of the CMB and (right column) matter
power spectra associated with a slope change at $k_0$ in $V(\phi)$,
Eq.~(\ref{slopechangelnP}) with (upper row) $\Delta A/A > 0$ and (lower row)
$\Delta A/A < 0$. The data and error bars for the CMB power spectrum are taken
from WMAP (squares), CBI \cite{cbi} (circles) and ACBAR \cite{acbar}
(triangles), and those for the matter one from SDSS \cite{sdss} respectively.
$k_0$ is fixed to be $0.05\mathrm{Mpc}^{-1}$.}
\end{figure}

First let us consider the case of the slope change\footnote{Our discussion here
is given in somewhat different context from Ref.~\cite{lps}.}. It is rather
straightforward to predict the CMB and matter power spectra. As shown in the
left panel of Figure~\ref{slopechange}, there is a power modulation on the
small scale for $\mathcal{P}(k)$, depending on the sign of the slope change.
Consequently, we expect that the CMB and matter power spectra would show the
corresponding change in power. In the upper panel of Figure~\ref{k_size}, we
set the position of the slope change to be $0.05\mathrm{Mpc}^{-1}$ and give a
few positive changes. On the left panel where the CMB spectrum is shown, we can
see clearly that there is an enhancement up to $l \sim 800$ when the slope
changes positively. The corresponding matter power spectrum is on the right
side. Compared with the large scale, the power on small $k$ is decreased
drastically as the slope change becomes bigger. In opposite, there is an
enhancement on small scales if the change is negative, as shown in the lower
row of Figure~\ref{k_size}. Taking matter power spectra into account, such an
enhancement is more prominent. For both cases, the diminishing oscillatory
behaviour is not noticeable in the CMB spectra, since the background CMB
spectrum is also oscillating. Rather, it is clearly shown in the matter power
spectra in the small $k$ region.

\begin{figure}[h]
\psfrag{LCDM}{$\Lambda$CDM}%
\psfrag{l(l+1)Cl/2p(mK2)}{$l(l + 1)C_l/2\pi (\mu K^2)$}%
\psfrag{l}{$l$}%
\psfrag{P(k)(Mpc)3}{$P(k)(\mathrm{Mpc})^3$}%
\psfrag{k(Mpc-1)}{$k (\mathrm{Mpc}^{-1})$}%
\psfrag{k_0=0.005}{$k_0 = 0.005$}%
\psfrag{k_0=0.01}{$k_0 = 0.01$}%
\psfrag{k_0=0.05}{$k_0 = 0.05$}%
\psfrag{k_0=0.1}{$k_0 = 0.1$}%
\begin{center}
\epsfig{file=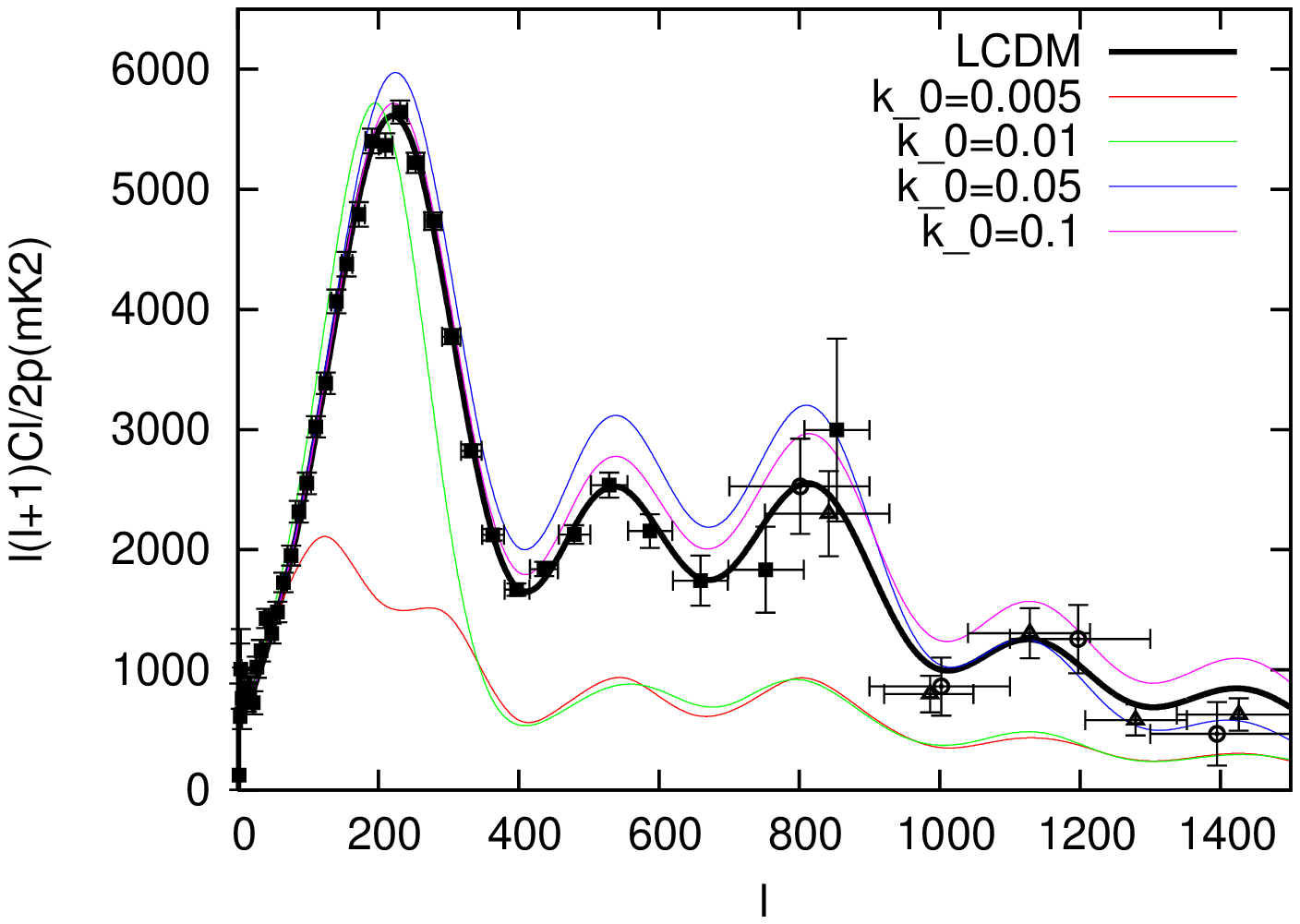, angle = -90, width = 8.1cm}
\epsfig{file=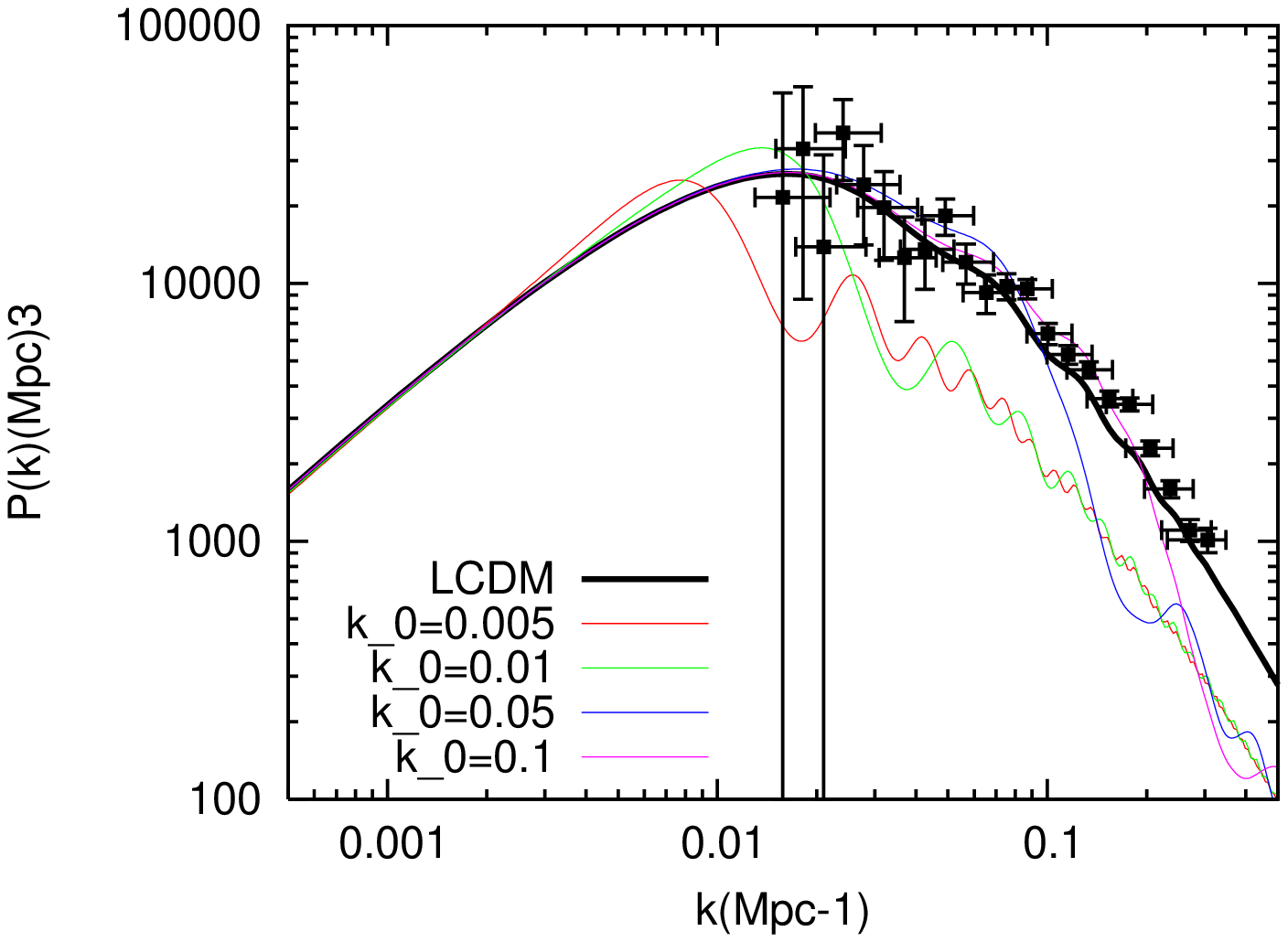, angle = -90, width = 8.1cm}
\end{center}
\begin{center}
\epsfig{file=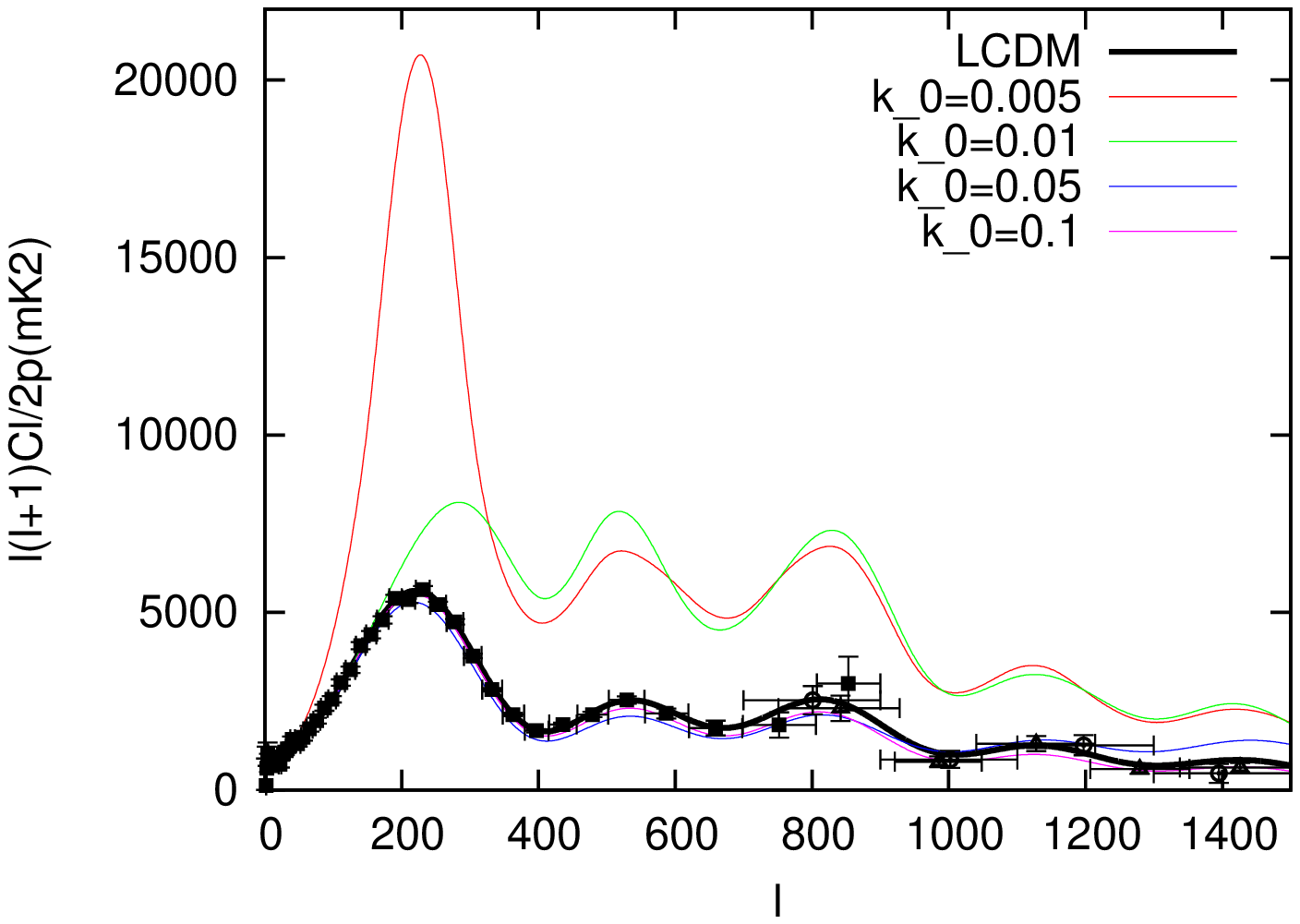, angle = -90, width = 8.1cm}
\epsfig{file=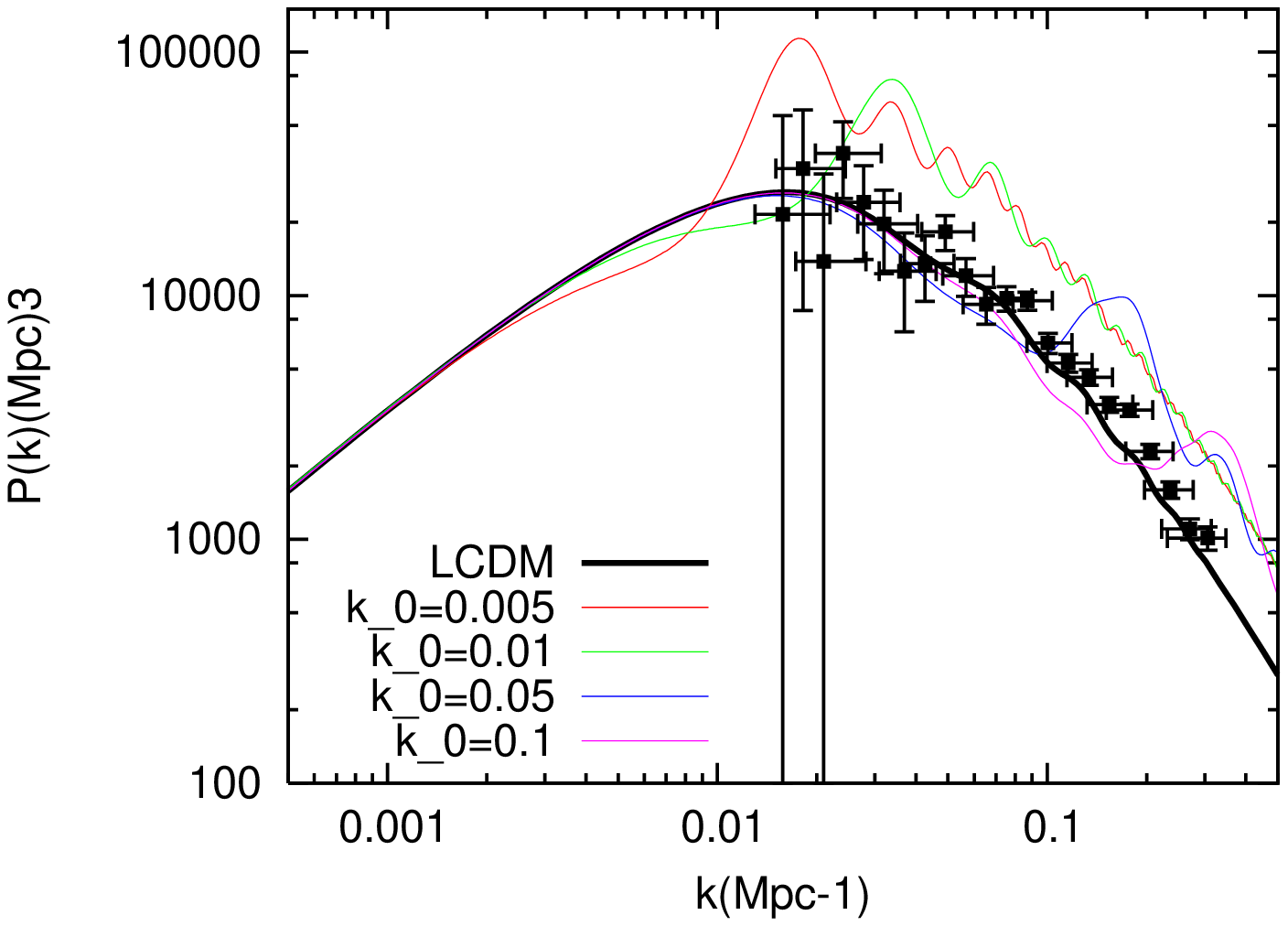, angle = -90, width = 8.1cm}
\end{center}
\caption{\label{kposition} (Left column) plots of the CMB and (right column)
matter power spectra with (upper row) positive and (lower row) negative slope
change, depending on the position of the slope change $k_0$. The size of the
slope change is set to $|\Delta A/A| = 0.5$.}
\end{figure}

We can also ask the dependence on the position of the slope change. If the slope
changes early enough, it would affect most of the observationally relevant
scales, and the CMB and matter power spectra should be very different from those
we actually observed. The effects of various positions of the slope change are
given in Figure~\ref{kposition}. As anticipated, when the slope change happens
before the largest observable scale, power spectra are largely enhanced or
suppressed on most of the scale.

\begin{figure}[h]
\psfrag{LCDM}{$\Lambda$CDM}%
\psfrag{l(l+1)Cl/2p(mK2)}{$l(l + 1)C_l/2\pi (\mu K^2)$}%
\psfrag{l}{$l$}%
\psfrag{P(k)(Mpc)3}{$P(k)(\mathrm{Mpc})^3$}%
\psfrag{k(Mpc-1)}{$k (\mathrm{Mpc}^{-1})$}%
\psfrag{B = 0.1}{$B = 0.1$}%
\psfrag{B = 0.5}{$B = 0.5$}%
\psfrag{B = 1.0}{$B = 1.0$}%
\psfrag{k_s=0.005}{$k_s = 0.005$}%
\psfrag{k_s=0.01}{$k_s = 0.01$}%
\psfrag{k_s=0.05}{$k_s = 0.05$}%
\psfrag{k_s=0.1}{$k_s = 0.1$}%
\begin{center}
\epsfig{file=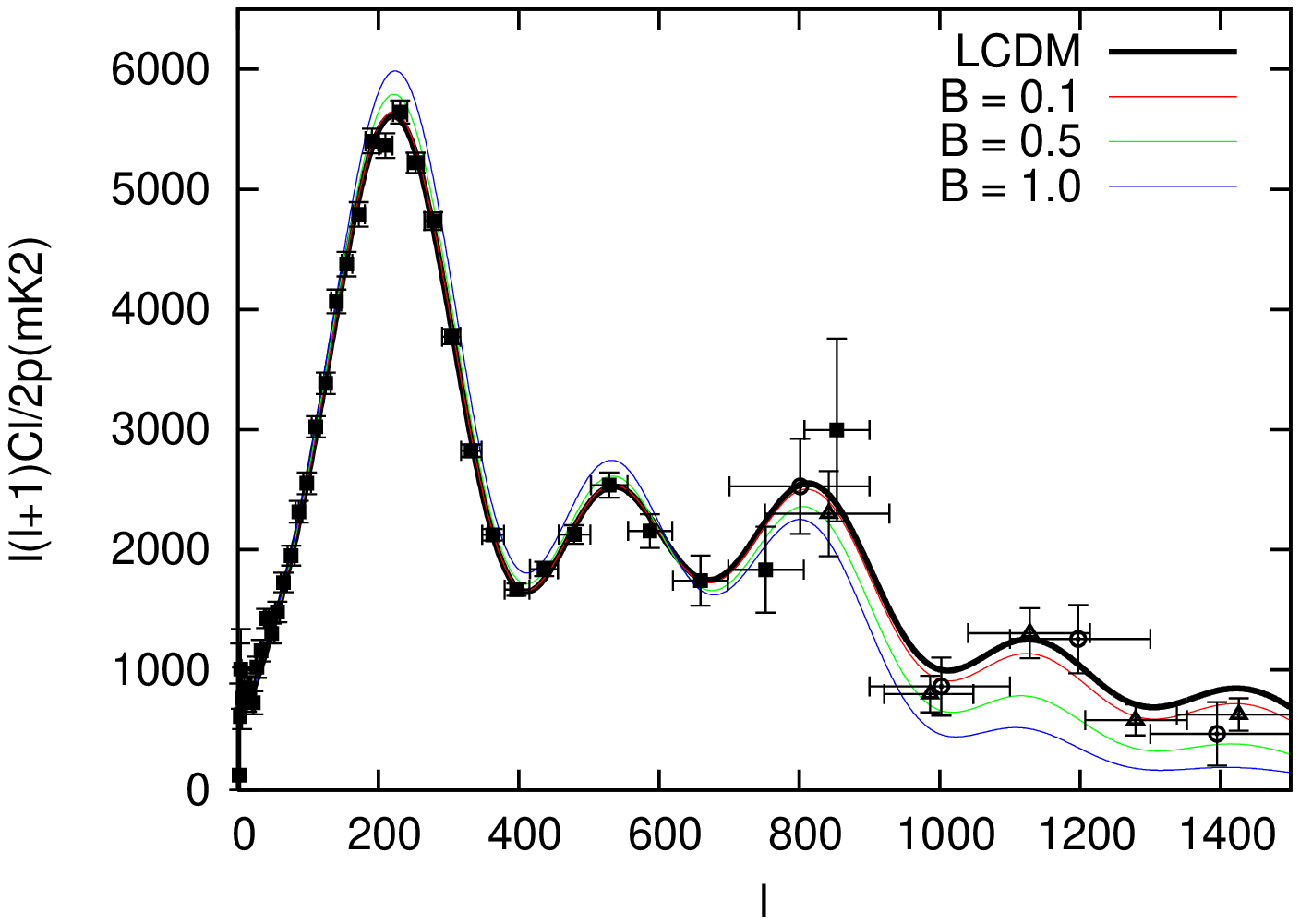, angle = -90, width = 8.1cm}
\epsfig{file=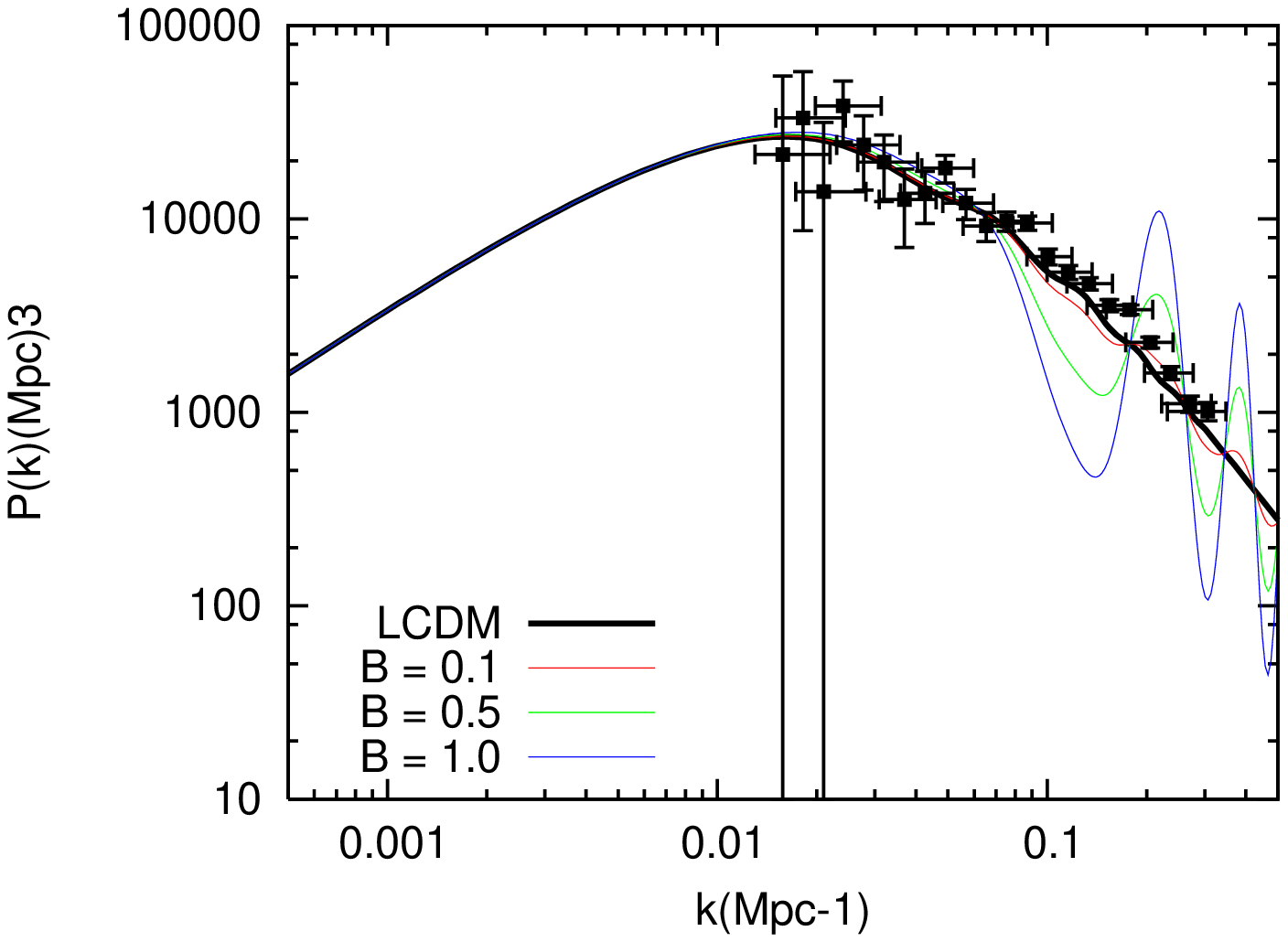, angle = -90, width = 8.1cm}
\end{center}
\begin{center}
\epsfig{file=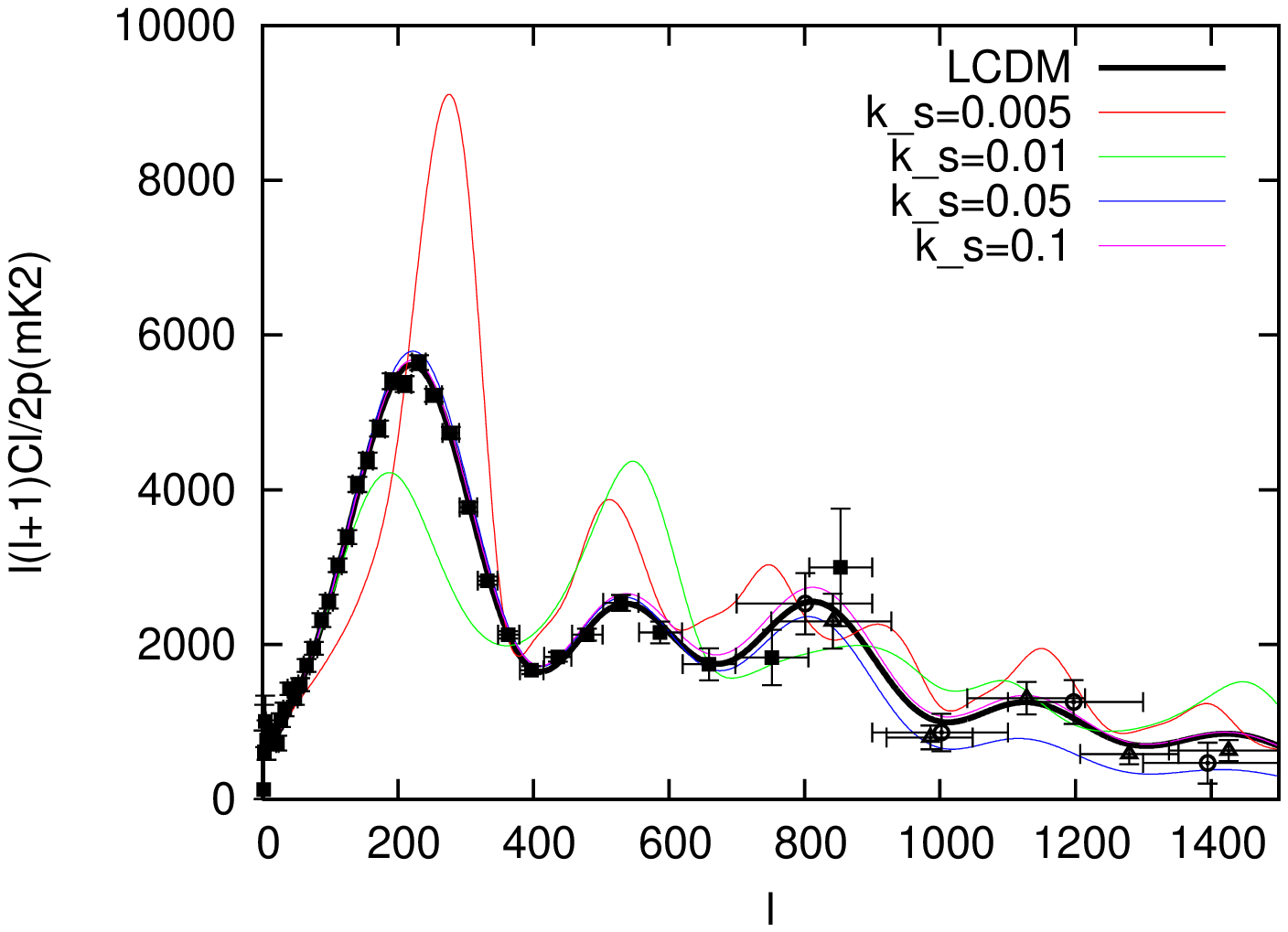, angle = -90, width = 8.1cm}
\epsfig{file=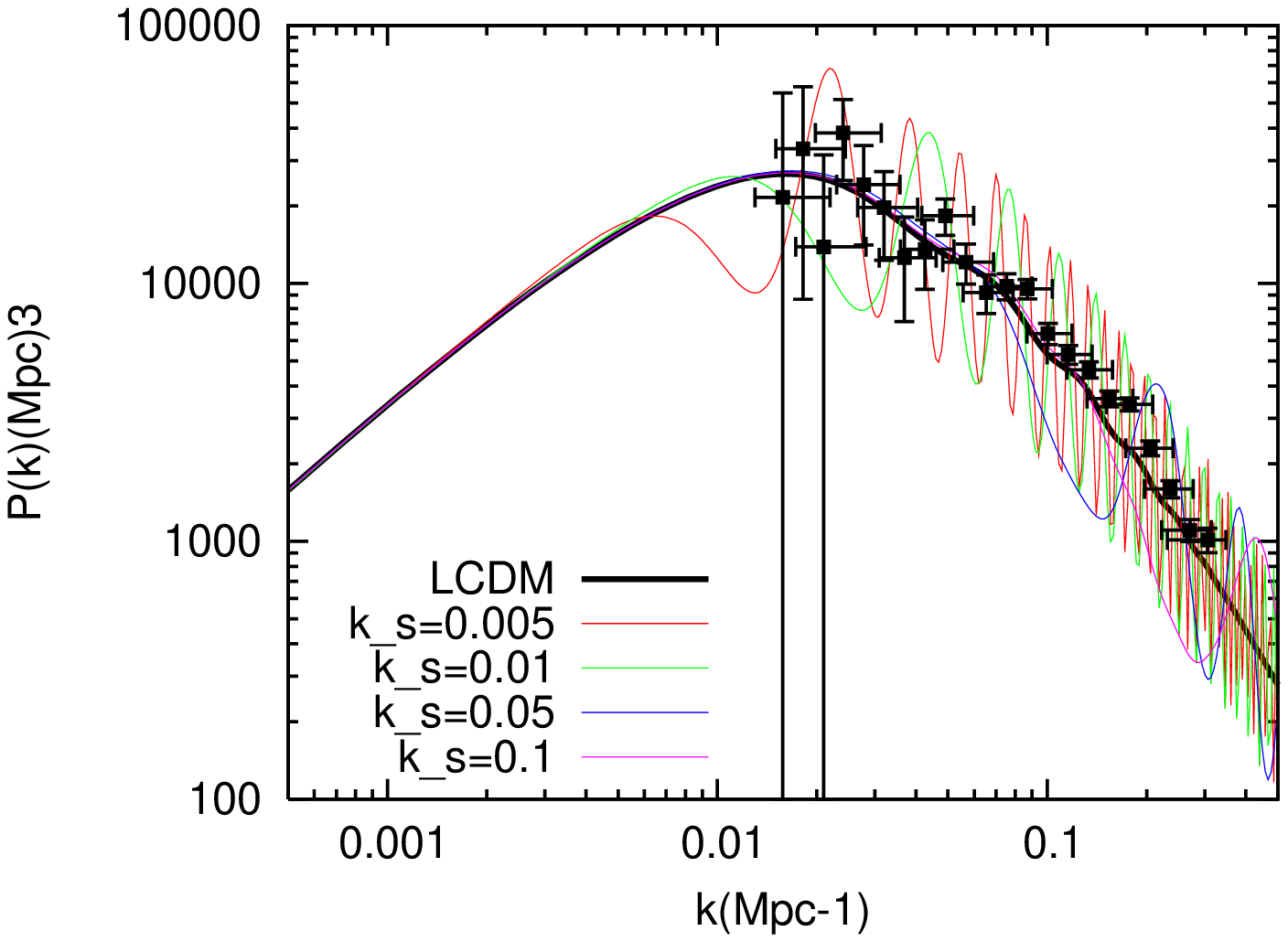, angle = -90, width = 8.1cm}
\end{center}
\caption{\label{step} (Left column) plots of the CMB and (right column) matter
power spectra corresponding to the sharp downward step in $V(\phi)$. In the
upper row $k_s$ is fixed at $0.05\mathrm{Mpc}^{-1}$, and in the lower row the
magnitude of the step is fixed to be $B = 0.5$ but the position $k_s$ is
varying.}
\end{figure}

Now we transfer our attention to the case of the step. Before we present the
results, let us pause for a moment and anticipate the associated spectra by
considering the form of $\mathcal{P}(k)$ in Figure~\ref{stepgraph}. It is clear
that around the scale $k_s$, which corresponds to the position of the step,
first we see that after a slight enhancement, there is a large suppression in
the power, and then oscillation. So, the CMB and matter power spectra should
first experience a suppression, and then oscillate. One thing here we should
bear in mind is the value of $k_s$. If $k_s$ is large, that is, if the step
appears quite lately, many of the relevant scales would have already crossed
the horizon and there would be no prominent difference. Conversely, when $k_s$
is small, i.e., most of the observationally important scales have gone through
the step, the oscillatory behaviour of $\mathcal{P}(k)$ would have been
obviously imprinted on the spectra we see today when the step is sharp enough.
In the upper row of Figure~\ref{step} $k_s$ is fixed to be
$0.05\mathrm{Mpc}^{-1}$ and the size of the step varies. In the CMB spectrum on
the left side it is shown that on small $l$ the power is slightly greater, but
on large $l$ it becomes smaller. The oscillatory feature is manifest in the
matter power spectrum in the right panel of Figure~\ref{step}. The dependence
of the CMB and matter power spectra on $k_s$ is depicted in the lower row of
Figure~\ref{step}. As expected, when $k_s$ is large, it is almost the same with
the background $\Lambda$CDM cosmology. But if $k_s$ is small enough to affect
the whole observable scales, the departure from the usual $\Lambda$CDM is very
large. Note that for the case of smoothed step, only the first few oscillations
in the matter power spectrum survive and the rest quickly diminish, as shown in
Figure~\ref{stepgraph}.

\subsection{Degeneracy}
\label{degeneracy}

As we have seen in the previous section, the distortions of the observable CMB
and matter power spectra are dependant on both the magnitude and the position
of the feature. Of these, the magnitude of the feature is tightly constrained
to be very small by cosmological observations on the CMB and galaxy
distribution. If the size were to be large enough to have been imprinted on the
observable spectra, we would already have noticed a large departure from scale
invariance on various scales. For example, the fractional change in the
inflaton potential amplitude of about $0.1\%$ for the smoothed hyperbolic
tangent step, i.e., $a \sim 0.001$, is known to cause sufficient sharp features
in the CMB spectrum \cite{ace,wmapfeature}. Yet, it seems more difficult to put
strong constraints on the position of the feature.

\begin{figure}[h]
\psfrag{LCDM}{$\Lambda$CDM}%
\psfrag{l(l+1)Cl/2p(mK2)}{$l(l + 1)C_l/2\pi (\mu K^2)$}%
\psfrag{l}{$l$}%
\psfrag{k_s=0.005}{$k_s = 0.005$}%
\psfrag{k_s=0.01}{$k_s = 0.01$}%
\psfrag{k_s=0.05}{$k_s = 0.05$}%
\psfrag{k_s=0.1}{$k_s = 0.1$}%
\begin{center}
\epsfig{file=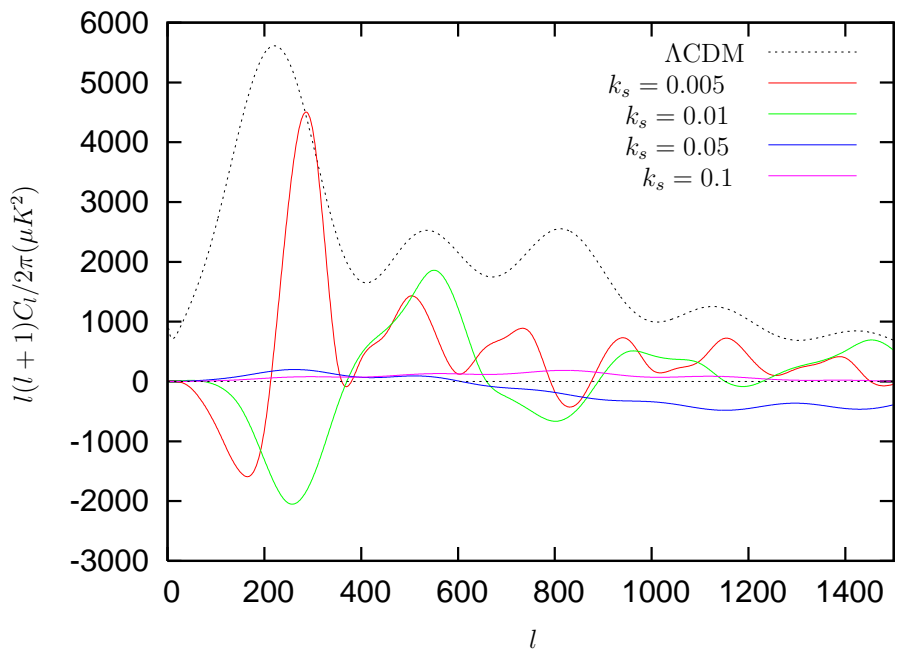,angle=-90,width=5 in} %
\caption{\label{steposc} The differences in the CMB power spectra between the
models with a step and the background $\Lambda$CDM, $C_l (\mbox{step models in
the lower row of Figure~\ref{step}}) - C_l (\Lambda\mathrm{CDM})$.}
\end{center}
\end{figure}

In the light of this, it seems not impossible to mimic some cosmological model
by introducing a feature to another one with different parameters\footnote{In
Ref.~\cite{bdrs}, it is shown that we can even imitate the concordant
$\Lambda$CDM with a model without $\Lambda$ ($\Omega_\Lambda = 0$) by relaxing
the hypothesis that the fluctuation spectrum can be described by a single power
law.}. That is, provided that the size is small enough, we may have strong
degeneracies between cosmological models depending on appropriate position of
the feature. In particular, by the CMB spectrum alone, we are possibly led to
determine the cosmological parameters incorrectly, since the oscillatory period
in the CMB spectrum might overlap with that of the feature. In
Figure~\ref{steposc}, we show the difference between the CMB spectrum of the
background $\Lambda$CDM and those of the models with steps at various
positions. We see that a step at some position slightly smaller than $k_s \sim
0.01\mathrm{Mpc}^{-1}$ gives a similar oscillatory cycle. Also, $k_s \lesssim
0.05\mathrm{Mpc}^{-1}$ seems to enhance the first peak slightly, but lower the
subsequent ones.

\begin{figure}[h]
\psfrag{l(l+1)Cl/2p(mK2)}{$l(l + 1)C_l/2\pi (\mu K^2)$}%
\psfrag{l}{$l$}%
\psfrag{P(k)(Mpc)3}{$P(k)(\mathrm{Mpc})^3$}%
\psfrag{k(Mpc-1)}{$k (\mathrm{Mpc}^{-1})$}%
\psfrag{LCDM}{$\Lambda$CDM}%
\psfrag{Step}{Step}%
\begin{center}
\epsfig{file=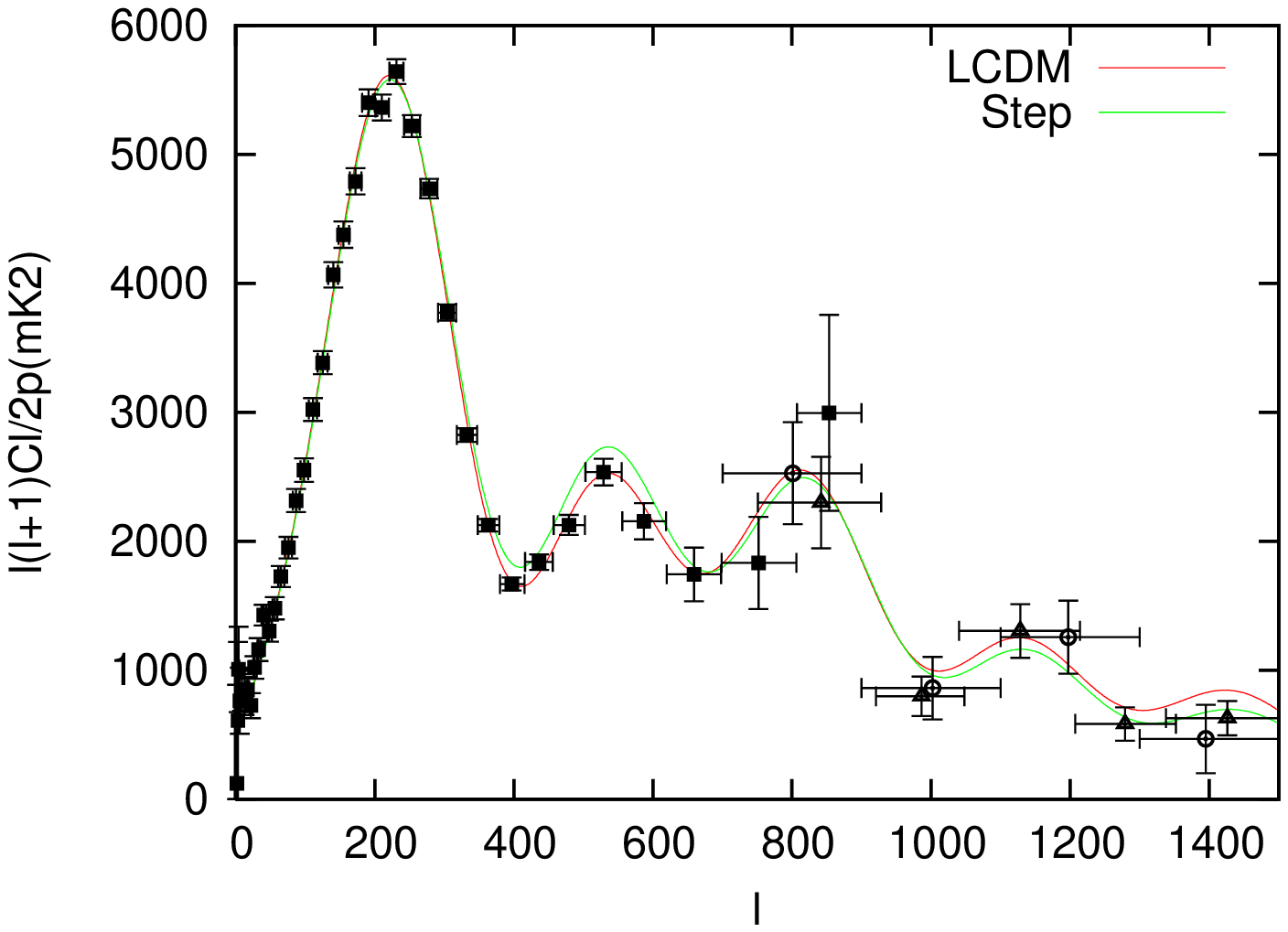, angle = -90, width = 8.1 cm}
\epsfig{file=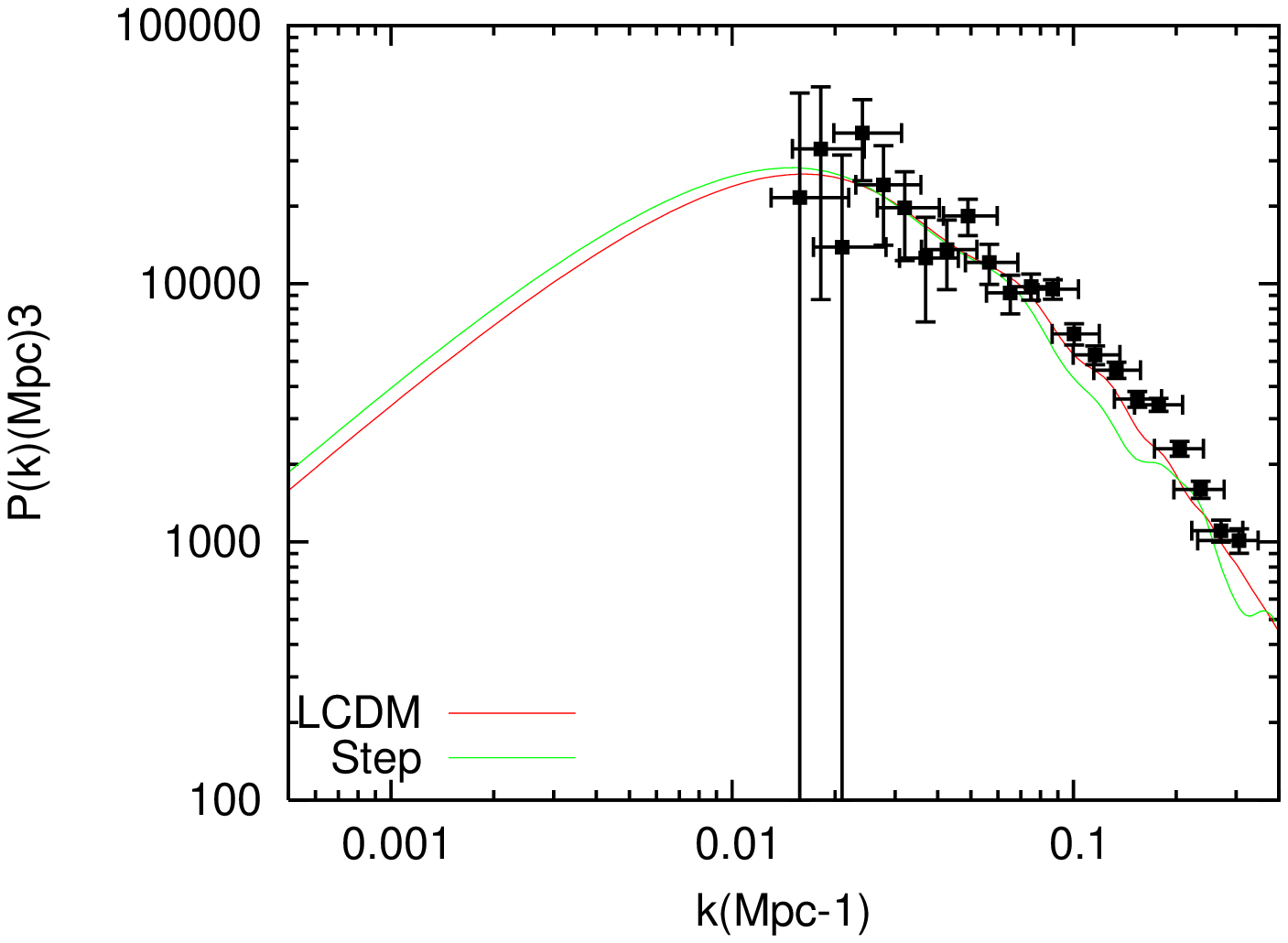, angle = -90, width = 8.1 cm}
\end{center}
\caption{\label{degeneracy1} Power spectra of the background $\Lambda$CDM and a
step model where $\Omega_\mathrm{B} = 0.040, \Omega_\mathrm{DM} = 0.200$ and $
\Omega_\Lambda = 0.760$. The first peaks of the two CMB spectra are nearly the
same.}
\end{figure}

For example, in Figure~\ref{degeneracy1}, we show the CMB and matter power
spectra of two different models. One model, with a scale invariant primordial
spectrum, is provided by the background $\Lambda$CDM, while the other has a
sharp downward step at $k_s = 0.01\mathrm{Mpc}^{-1}$ with $B = 0.1$, where the
cosmological parameters are set to $\Omega_\mathrm{B} = 0.040,
\Omega_\mathrm{DM} = 0.200$ and $\Omega_\Lambda = 0.760$. In this case, the
position and the height of the first peak are very similar for the two CMB
spectra, making them not easily distinguishable. However, in the matter power
spectra we can note an oscillation in the power in the small $k$ region,
leaving some scope for discrimination.

\begin{figure}[h]
\psfrag{LCDM}{$\Lambda$CDM}%
\psfrag{l(l+1)Cl/2p(mK2)}{$l(l + 1)C_l/2\pi (\mu K^2)$}%
\psfrag{l}{$l$}%
\psfrag{P(k)(Mpc)3}{$P(k)(\mathrm{Mpc})^3$}%
\psfrag{k(Mpc-1)}{$k (\mathrm{Mpc}^{-1})$}%
\psfrag{Step}{Step}%
\begin{center}
\epsfig{file=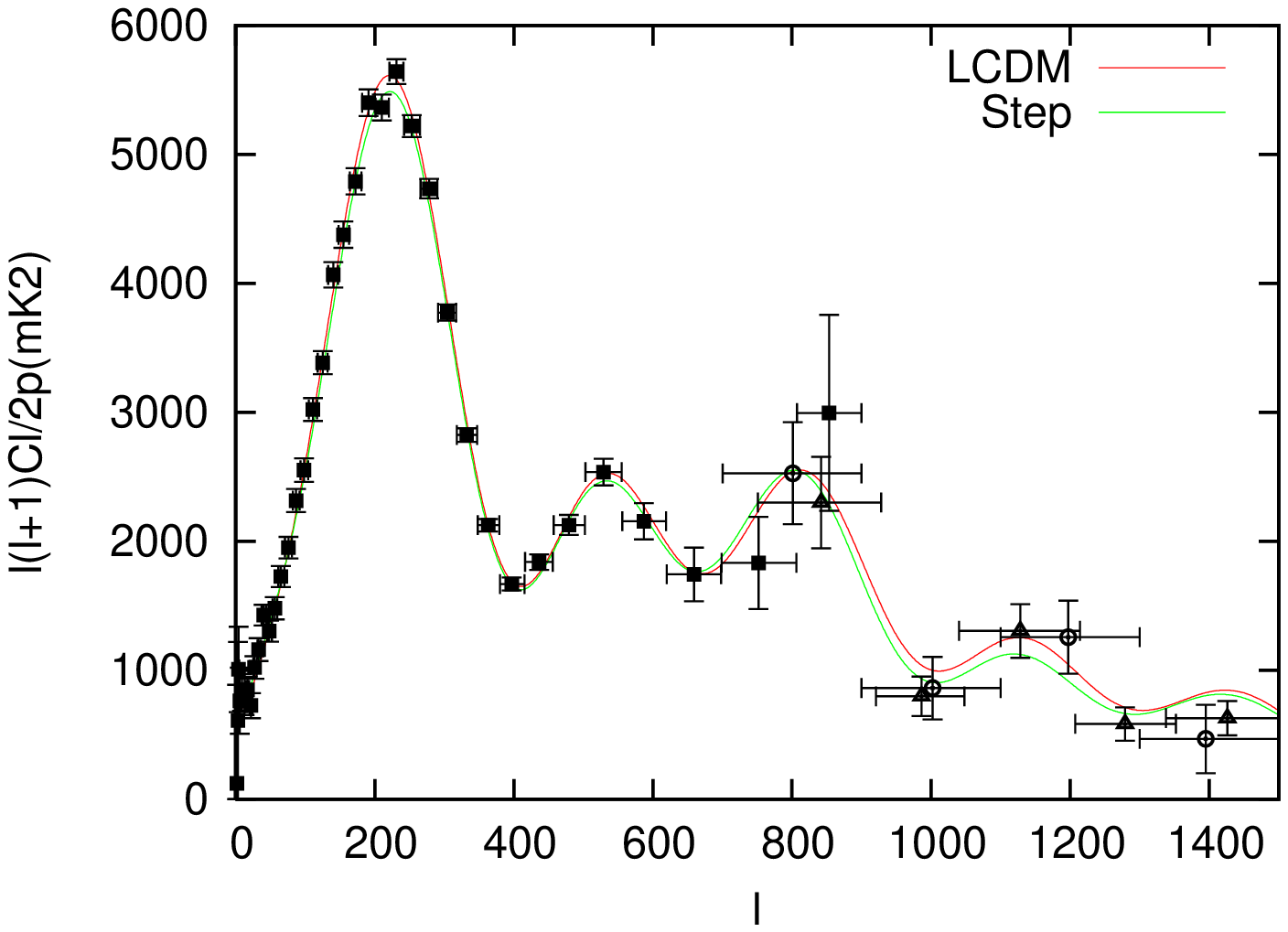, angle = -90, width = 8.1 cm}
\epsfig{file=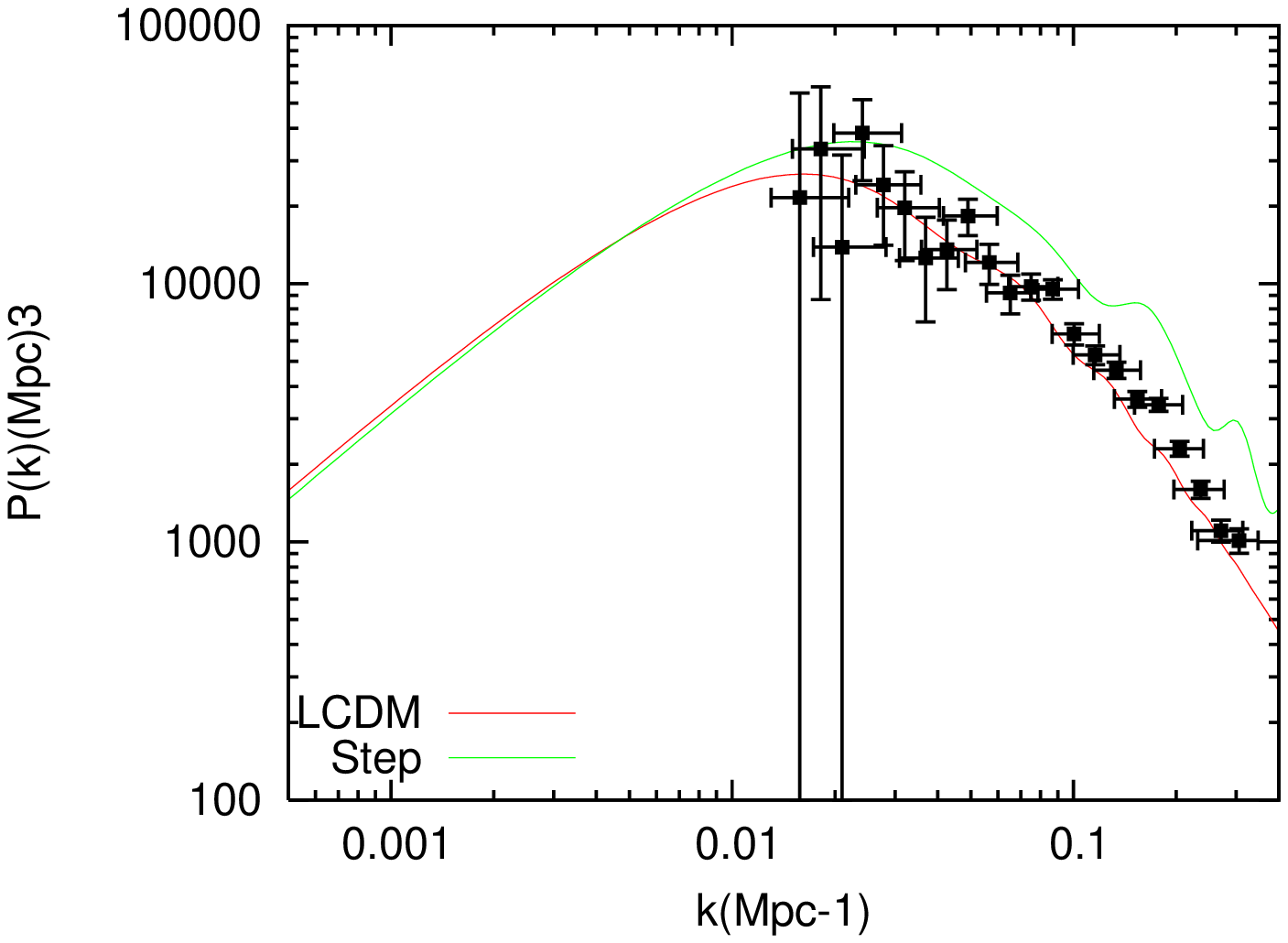, angle = -90, width = 8.1 cm}
\end{center}
\caption{\label{degeneracy2} Power spectra of the background $\Lambda$CDM and a
step model where $\Omega_\mathrm{B} = 0.060, \Omega_\mathrm{DM} = 0.390,
\Omega_\Lambda = 0.550$ and $h = 0.62$. Both the two CMB power spectra seem well
within the error bars in the CMB spectrum observation. Can we distinguish these
two models with the CMB observation alone?}
\end{figure}

Another example is given in Figure~\ref{degeneracy2}, where again the
background $\Lambda$CDM is shown for comparison. Here we take $B = 0.1$ and
$k_s = 0.04\mathrm{Mpc}^{-1}$, and the parameters $\Omega_\mathrm{B} = 0.060,
\Omega_\mathrm{DM} = 0.390, \Omega_\Lambda = 0.550$ and $h = 0.62$. While the
cosmological parameters are quite different, the CMB peaks and troughs roughly
overlap in this example. It seems that even if we could measure the CMB
spectrum with very high accuracy by some experiments, e.g., using the planned
Planck surveyor, we should be very cautious to abandon any featured primordial
spectrum and choose the precisely flat, scale invariant one confidently.

Therefore it is very likely that with the CMB observation alone, where the
usually assumed scale invariant $\mathcal{P}(k)$ of the standard $\Lambda$CDM
and the scale dependent $\mathcal{P}(k)$ from the inflaton potential with a
step generate very similar CMB power spectra despite the different cosmological
parameters, we are led to confusion for the parameter estimation. Such
degeneracies would only be broken by combining some other observations, e.g., a
matter power spectrum, as shown in Figure~\ref{degeneracy2}. But when the step
is not infinitely sharp but mild, which is supposed to be a more realistic
case, the oscillation quickly dies away, except a first couple of peaks, back
to the background value. This should make discrimination difficult in the
matter power spectrum observations as well.

\section{Conclusions}
\label{seccon}

In many extensions of the standard model of particle physics, e.g.,
supersymmetry and string theory, the existence of many scalar fields is
predicted. During inflation which such theories would have described correctly,
more than one scalar field may have been relevant. In this regards, it is
meaningful to investigate the effect of inflaton potential with some feature
which is produced generically by spontaneous symmetry breaking of another
scalar field coupled to inflaton. The corresponding power spectrum of curvature
perturbations $\mathcal{P}(k)$ generally deviates from the usually adopted
flat, scale invariant one.

In this paper, we have considered the primordial power spectrum of curvature
perturbations which arises from some potential with a feature due to
spontaneous symmetry breaking of another scalar field coupled to inflaton
field. On the basis of the general slow-roll approximation, we can proceed in a
systematic and organised manner with the calculation of the curvature power
spectrum $\mathcal{P}(k)$, and indeed in Section~\ref{secps} we have obtained
fully analytic results of $\mathcal{P}(k)$ in various situations. Generally,
such a feature leads to the breaking of scale invariance of $\mathcal{P}(k)$,
which is given as an oscillatory behaviour in $\mathcal{P}(k)$ on scales
smaller than those associated with the position of the feature, accompanied
with an overall modulation of the power when $V(\phi)$ becomes flatter or
steeper. Given such a power spectrum, we also have obtained the associated CMB
and matter power spectra which we can actually observe. When the magnitude of
the feature is large and its position corresponds to large scale, both the
oscillation and power modulation imprinted in $\mathcal{P}(k)$ are noticeable
in the CMB and matter power spectra.

One expectation we could draw from the above results is that it might be
possible to mimic some cosmological model with a scale invariant primordial
power spectrum $\mathcal{P}(k)$ with a scale dependent $\mathcal{P}(k)$
resulting from some featured inflaton potential in spite of different
cosmological parameters such as $\Omega_\mathrm{B}, \Omega_\mathrm{DM},
\Omega_\Lambda$ and $h$. Such a degeneracy is especially strong for the CMB
power spectrum, since the oscillatory period of $\mathcal{P}(k)$ may overlap
with that of the CMB peaks, depending on the magnitude and the position of the
feature in $V(\phi)$. In this regard, we believe that close investigations of
independent observations, e.g., the matter power spectrum or supernovae, would
remove our potential confusion as regards the cosmological parameter estimates.
It might also provide interesting theoretical possibilities for the particle
physics responsible for the generation of the feature in $V(\phi)$. Tight
observational bounds on the magnitude and the position of the feature would
constrain the particle physics model behind the inflaton potential as well,
shedding some light on our understanding of the universe.

\subsection*{Acknowledgements}

I thank Kiwoon Choi, Jai-chan Hwang, Donghui Jeong, Kang Young Lee, Carlos
Mu\~noz, Subir Sarkar, Misao Sasaki and Ewan Stewart for many helpful
discussions and important suggestions. Especially I am indebted to Richard
Easther for invaluable comments on the CMB spectrum. This work was supported in
part by ARCSEC funded by the KOSEF and the Korean Ministry of Science and the
KRF grant PBRG 2002-070-C00022.

\end{document}